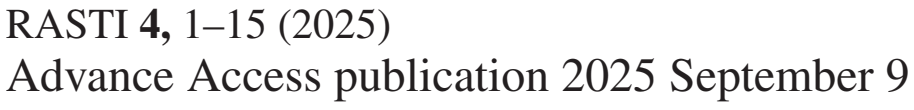

# MADVORO : parallel construction of Voronoi diagrams in distributed memory systems

Maor Mizrachi ,[1]★ Barak Raveh[1] and Elad Steinberg [2]
[1]*School of Computer Science and Engineering, The Hebrew University, 9190401 Jerusalem, Israel*
[2]*Racah Institute of Physics, The Hebrew University, 9190401 Jerusalem, Israel*

## ABSTRACT

Voronoi diagrams are essential geometrical structures with numerous applications, particularly astrophysics-driven finite volume methods. While serial algorithms for constructing these entities are well-established, parallel construction remains challenging. This is especially true in distributed memory systems, where each host manages only a subset of the input points. This process requires redistributing points across hosts and accurately computing the corresponding Voronoi cells. In this paper, we introduce a new distributed construction algorithm, which is implemented in our open-source C++ 3D Voronoi construction framework. Our approach leverages Delaunay triangulation as an intermediate step, which is then transformed into a Voronoi diagram. We introduce the algorithms we implemented for the precise construction and our load-balancing approach and compare the running time with other state-of-the-art frameworks. MADVORO is a versatile tool that can be applied in various scientific domains, such as mesh decomposition, computational physics, chemistry, and machine learning.

## 1 INTRODUCTION

The Voronoi diagram is an elementary geometry structure. It is widely used for many purposes across various disciplines, including mathematics, computer science, physical sciences, and even health and social sciences. For example, in computer science, Voronoi diagrams play a crucial role in Lloyd's algorithm, which underpins methods such as *k*-means clustering. In physics, Voronoi diagrams can be used to design a mesh decomposition in FVM (Finite-Volume Method) simulations, for example, as described in Yalinewich, Steinberg & Sari (2015) and Springel (2010). In chemistry, they can be used to analyse protein structures.

In this paper, we introduce a novel distributed-memory framework for Voronoi diagram construction, originally developed for the RICH hydrodynamic simulation (Yalinewich et al. 2015). Our framework is specifically designed for astrophysical simulations, where consecutive mesh reconstructions are required, and it excels in handling complex meshes commonly encountered in this domain, as we will demonstrate. Due to memory and computational constraints, the construction must be performed in parallel. Each processor is responsible for a subset of the domain, constructing the Voronoi diagram locally while ensuring proper connectivity with remote cells managed by other processors. In Voronoi-based simulations, the mesh must be reconstructed at each time-step, making mesh generation a critical bottleneck. By optimizing this process, our method significantly reduces construction time, thereby improving overall computational efficiency. MADVORO is an object-oriented C++ library designed to efficiently construct and analyse 3D Voronoi diagrams. Its API provides structured access to a wide range of topological and geometric properties, as follows: the set of faces and vertices defining each cell, the connectivity between neighbouring cells, the normal vectors and centroids of individual faces, the centre of mass of each cell, as well as precise measurements such as face areas and cell volumes.

The rest of the paper is organized as follows. Section 2 deals with Voronoi diagrams construction and the load balancing problem, an essential problem related to distributed construction. Section 3 provides an overview of Voronoi diagrams, Delaunay triangulations, and their construction methods. In Section 4, we introduce the algorithms implemented by Springel's in Springel (2010). In Section 5, we describe our load-balancing approach. In Section 6, we present our algorithm, which ensures a valid Delaunay triangulation construction in distributed memory, then to be translated into a Voronoi diagram. Section 7 evaluates the methods compared to other construction methods. Finally, Section 8 concludes the paper and outlines directions for future research.

## 2 PREVIOUS STUDIES

### 2.1 Previous work on load balancing

To compute the Voronoi diagram in distributed memory or even in parallel memory, one has to consider re-distributing the generating points to alleviate the construction process. This process resembles mesh decomposition in physical simulations, where the space is broken down to a mesh, later to be partitioned into zones, each under the responsibility of a single processor. A key objective is to minimize

★ E-mail: maormiz@cs.huji.ac.il

communication overhead while ensuring a balanced computational load across processors.

Our load balancing problem is equivalent to the graph partitioning problem, where the graph's vertices are the points or Voronoi cells, and the edges represent a shared face. Unfortunately, the graph partitioning problem is known as NP-hard to solve, and many approximations and heuristics attempt to address its solution. Many of these methods rely on multilevel graph partitioning, where the graph is recursively coarsened, partitioned at a simplified level, and then refined to achieve a full partition. These methods are called multilevel graphs and differ by refining rules and partition steps. Chen, Saad & Zhang (2022) describes many of those techniques. A well-known framework is PARMETIS,[1] which is a distributed graph partitioning framework, based on the serial METIS[2] code (Karypis & Kumar 1998). Other common frameworks are JOSTLE[3] (Walshaw & Cross 2007), SCOTCH,[4] and ZOLTAN.[5] They are all compared in Bokhare & Metkewar (2019), along with other frameworks.

Steinberg et al. (2015) introduced a load-balancing approach specifically designed for unstructured meshes, focusing on Voronoi diagrams. In their method, given a set of $P$ processors, they construct a secondary Voronoi mesh which is then used to assign spatial subregions to different processors. This technique has gained prominence in successive Voronoi constructions within astrophysical simulations, where the processor-based tessellation dynamically adjusts itself using heuristic methods.

Orthogonal Recursive Bisection (ORB) is also a common method for mesh decomposition, especially in SPH simulations. In this method, we construct a spatial data structure called KD-tree, a binary tree used to contain points. The tree nodes represent boxes. Each time, two children are created by cutting the box along one axis, by pre-defined rules, until reaching a constant number of points inside a box. The latter explains why the method is referred to as recursive bisection. Eventually, leaf boxes are assigned to different processors. This method is employed in many physics code, such as PKDGRAV3 (Potter, Stadel & Teyssier 2017), Gasoline (Wadsley, Stadel & Quinn 2004), and Morozov & Peterka (2016). In fact, KD-trees are used for multiple algorithms as well, as described and demonstrated by Morozov & Peterka (2016). Spectral graph partitioning (Pothen, Simon & Liou 1990) is a well-established theoretical technique that utilizes the spectral properties of a graph to generate partitions, typically aiming for a high ratio of internal to external edges. However, these methods are computationally expensive and often challenging – or even infeasible – to implement efficiently in distributed-memory systems. Curve-based load balancing, where the space is partitioned according to the behaviour of a 1D curve (see Subsection 5.1), has been studied theoretically and experimentally. A commonly used curve is the Hilbert curve, whose locality preserving properties were first rigorously defined in Gotsman & Lindenbaum (1996). Further analyses of this method can be found in Moon et al. (2001) and Bauman (2006). Harlacher et al. (2012) analysed the performance of curve-based load balancing for distributed meshes.

Borrell et al. (2018) compared the curve-based load balancing to other heuristics. Other studies were done by Sasidharan, Dennis & Snir (2015) (2D case) and Filipiak (2013). Further details can be found in Mizrachi et al. (2024).

[1]https://github.com/KarypisLab/ParMETIS
[2]https://github.com/KarypisLab/METIS
[3]https://chriswalshaw.co.uk/jostle/
[4]https://www.labri.fr/perso/pelegrin/scotch/
[5]https://sandialabs.github.io/Zoltan/

## 2.2 Voronoi and Delaunay construction frameworks and methods

Due to the widespread interest in Voronoi diagrams, their construction has been extensively studied, and several open-source frameworks are available.

VORO++[6] (Rycroft 2009) provides a C++ implementation for constructing 3D Voronoi tessellations. Another well-known tool is QVORONOI, part of the QHULL[7] software, which computes Voronoi diagrams by first constructing the Delaunay triangulation using a projection-based algorithm. Additionally, the Computational Geometry Algorithms Library (CGAL) offers an implementation for both Voronoi and Delaunay constructions (Alliez et al. 2010).

Lo Lo (2012) also introduced a parallel framework for constructing Delaunay triangulations in two and three dimensions. Their approach follows a common strategy used by other frameworks: partitioning the space into zones – often using a Cartesian grid or a KD-tree – constructing the triangulation locally within each zone, and then incorporating points from neighbouring zones. This point exchange process can be performed using various techniques. Duffell et al. (Duffell & MacFadyen 2011) introduced a serial 2D code based on Voronoi diagram decomposition. Their construction method relies on edge flipping; however, this approach does not generalize trivially to three dimensions. A distributed framework for constructing Delaunay or Voronoi diagrams must overcome a critical challenge: correctly identifying and retrieving the boundaries of all cells, including those adjacent to cells managed by other processors. These borders are defined by *ghost cells* or *ghost points*, which are points from other processors that will later be used to construct the local diagram. The main challenge is identifying the necessary ghost points and efficiently partitioning the points among processors to optimize the overall construction process. Once the ghost points are known, constructing the Voronoi diagram or Delaunay triangulation locally becomes straightforward, often utilizing the libraries previously discussed.

An early distributed Voronoi construction was introduced as open-source software by TESS2,[8] following the method described in Morozov & Peterka (2016). The KD-tree used to make a mesh decomposition (as described earlier), is then used to build the Delaunay triangulation. Local points ask to import points from neighbouring blocks (and, if necessary, neighbours of neighbours, and so on) until the local Delaunay triangulation is completed. A similar algorithm will be presented later.

Peterka et al. (Peterka, Morozov & Phillips 2014) and González (PARAVT; Gonzalez 2016) proposed a similar technique and also proved the correctness of the algorithm. PARAVT offers two distinct domain decomposition approaches; however, like the KD-Tree method, it partitions the space into rectangular blocks that are subsequently assigned to processors. Wu et al. (2023) presented PARVORO++ – a distributed Voronoi tessellation code. They provide two frameworks: one for automatic ghost-point search and another for manual search. In the manual approach, the user specifies a radius within which all ghost points for a given point are guaranteed to be located, thus reducing the search effort. They evaluated the tessellation time and strong scaling performance of their framework.

The authors of Singh, Byrohl & Nelson (2024) introduced VOTESS, a 3D distributed Voronoi diagram construction framework. Unlike other frameworks, including our own, this paper presents a code with a construction algorithm specifically designed for execution on GPUs. The algorithm works by independently determining the list of nearest neighbours for each cell, and then constructing the cell by clipping a box based on the bisections (Voronoi faces). The framework has demonstrated successful acceleration on GPUs when handling sufficiently large point data sets. However, it should be noted that the presented framework is a multithreaded code designed for parallel systems with shared memory.

[6]https://github.com/chr1shr/voro
[7]http://www.qhull.org/html/qvoronoi.htm
[8]https://github.com/diatomic/tess2

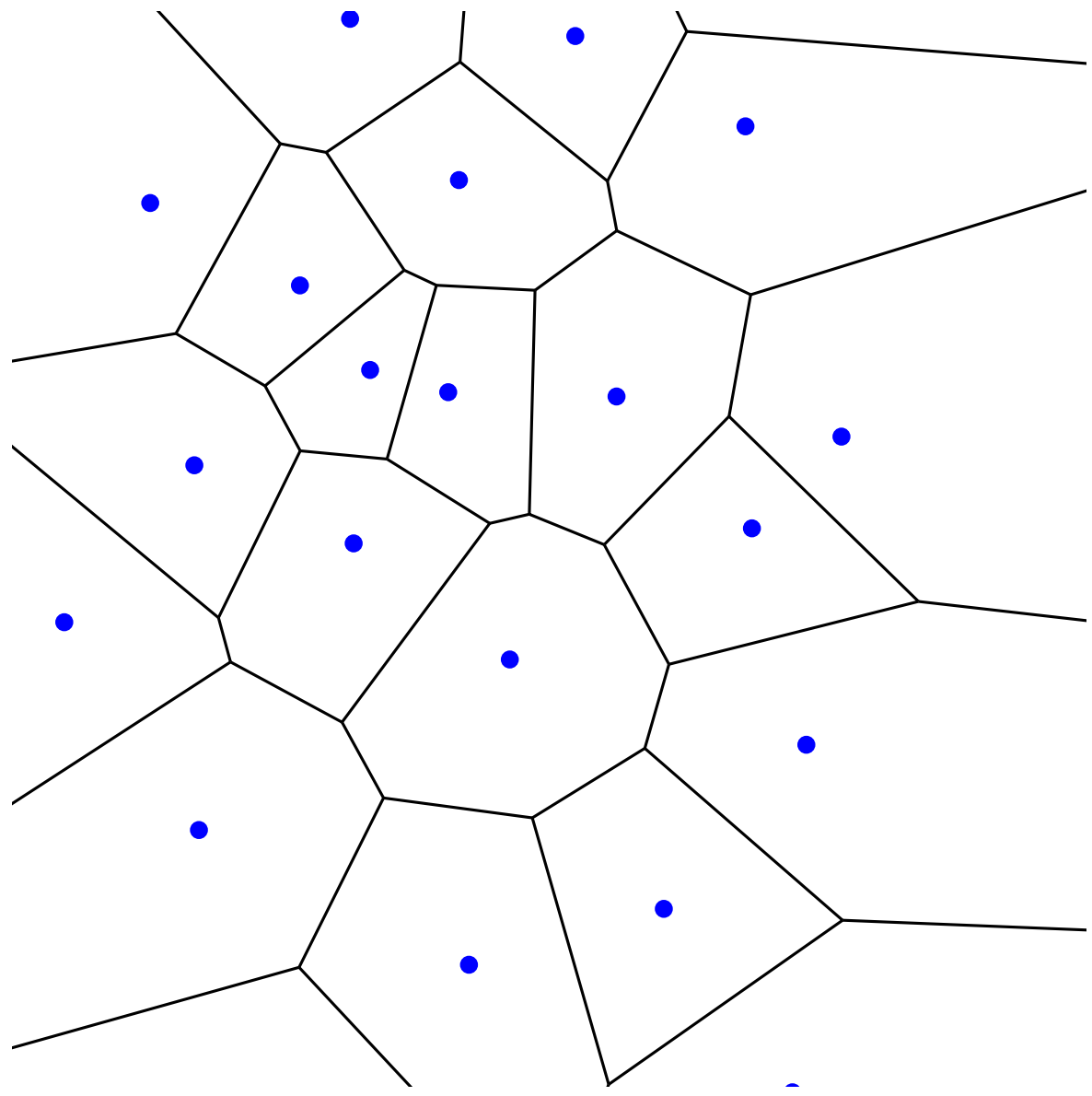

**Figure 1.** A Voronoi diagram of the set of 2D points (the border cells are infinite, and the figure is clipped).

Springel (2010), in his astrophysical code AREPO, which uses Voronoi diagrams as a spatial discretization method, proposed two algorithms for identifying ghost points and constructing the Delaunay triangulation (which is later converted into a Voronoi diagram). Unlike other KD-Tree implementations, Springel employs the Peano-Hilbert curve for space decomposition. In this paper, we present Springel's algorithms and enhance them to address challenges in more complex scenarios.

## 3 VORONOI DIAGRAMS AND DELAUNAY TRIANGULATIONS

### 3.1 Voronoi diagrams

#### *3.1.1 Definition*

A *Voronoi diagram* of a set of points $S \subseteq \mathbb{R}^n$ is defined as the partition of $\mathbb{R}^n$ into cells, each is the subspace of all the points in $\mathbb{R}^n$ closer to a certain point $p \in S$ than to other points of $S$. The points of $S$ are called the *generating points*. A point induces a cell and vice versa, and therefore, we interchangeably refer to a generating point and the cell induced by this point as the same. In cases where a point is equidistant from two points in $S$, it is arbitrarily assigned to a cell to keep the Voronoi diagram a valid partition.

In this way, each point in $S$ corresponds to a Voronoi cell, and vice versa.

The cells of a Voronoi diagram are unique for a given set of mesh-generating points. An example of a Voronoi diagram is portrayed in Fig. 1.

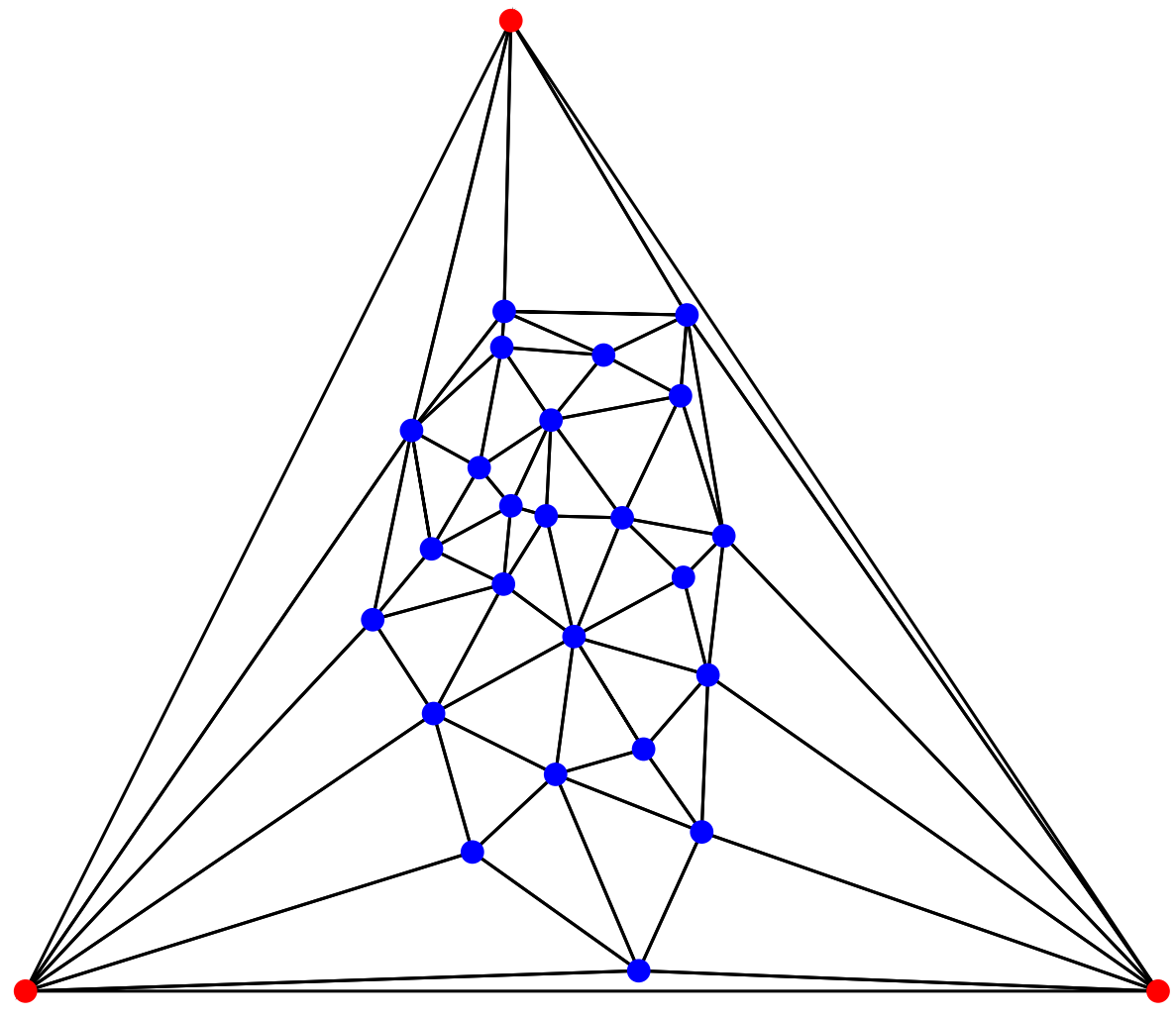

**Figure 2.** Delaunay triangulation of points (in blue). Red points are the bounding triangle's points.

#### *3.1.2 Construction of the Voronoi diagram*

The importance of the Voronoi diagram raises questions about how to find the Voronoi diagram of a given set of points $S$. By 'finding the Voronoi diagram,' we refer to identifying the neighbouring cells of each cell and determining the vertices of these cells. Since each cell corresponds to a point, this is equivalent to determining which points in $S$ are neighbours.

Multiple methods exist to construct the Voronoi tessellation of a given set of points. These methods are detailed and discussed in Watson (1993). In this paper, we focus on one such method, which involves an equivalent construction of a geometric structure known as the Delaunay triangulation.

### 3.2 Delaunay triangulations

#### *3.2.1 Definition*

A *triangulation* of a set of points (often called *sites* in this context) $S$ partitions the space into triangles,[9] whose vertices are all in $S$. A Delaunay triangulation of a set of points $S \subseteq \mathbb{R}^n$ is a special triangulation of $S$. It is a triangulation that follows the *Delaunay property*, or the *Empty Circumcircle Property*. The empty circumcircle property states that when examining the circumcircle of a triangle $T$, assuming $T$ is defined by the vertices $v_0, v_1, v_2 \in S$, there are no other points from $S$ inside the circumcircle.

Boundary points may not have corresponding points to form some triangles. Therefore, it is common to enclose the points within a large triangle that contains all the points and construct the Delaunay triangulation using these additional points as well.

An example is shown in Fig. 2.

#### *3.2.2 Delaunay–Voronoi duality*

A well-established result is that the Delaunay Triangulation of a set of points $A$ and the Voronoi Diagram of the same set of points are conjugate and equivalent in computation. The centres of the circumcircles of the triangles constitute the vertices of the Voronoi cells. In addition, assuming a set of generating points $S$, then if $a$ and $b$ are both neighbouring cells in $S$'s Voronoi diagram, the edge $(a, b)$ exists in $S$'s Delaunay triangulation.

[9]The definition of a triangle in dimensions higher than 2 is defined more often as *simplex*.

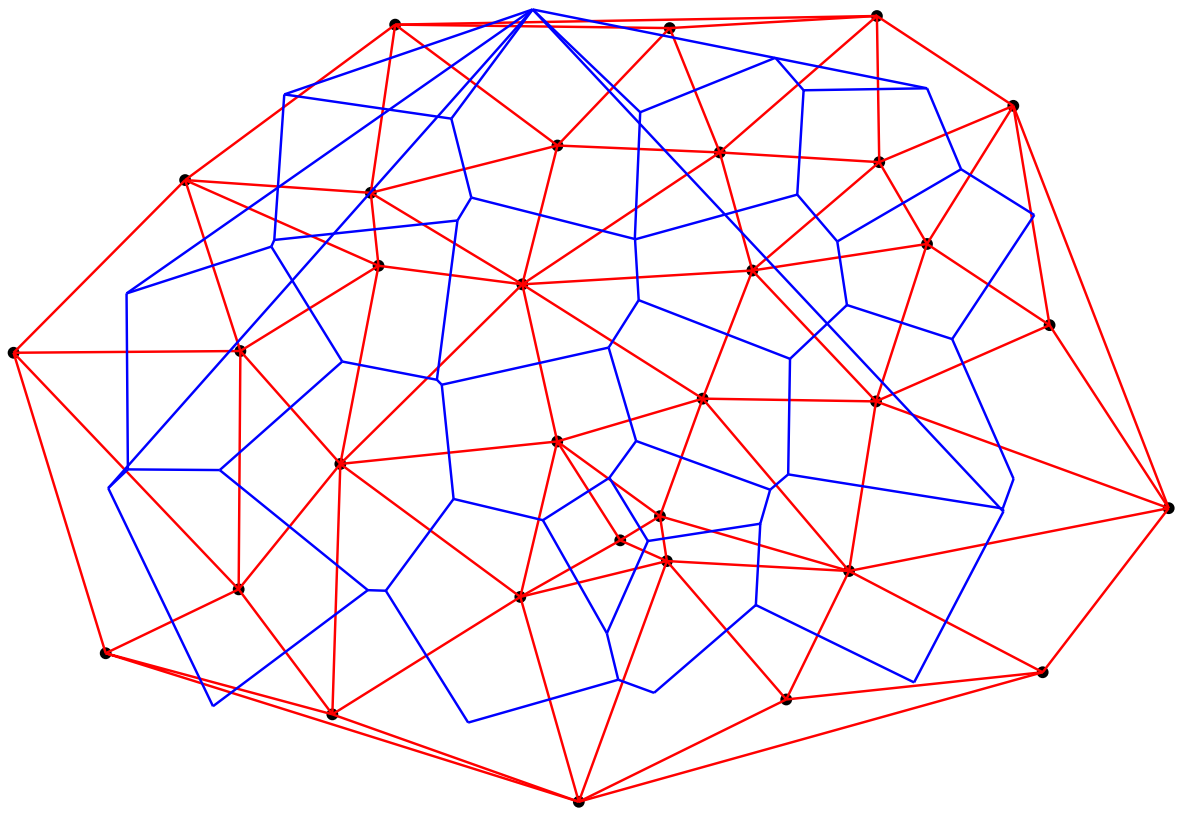

**Figure 3.** The duality between the Delaunay triangulation of a set of points (in red) and its Voronoi diagram (in blue). For example, the centre of the circumcircle of a triangle in the Delaunay triangle is a vertex of the Voronoi diagram.

The duality is more profound, as a one-to-one dual mapping can be done from $k$-dimensional objects in the Voronoi diagram to $(n-k)$-dimensional objects in the Delaunay triangulation. Fig. 3 shows an example of this duality. As previously noted, our approach to building the Voronoi diagram is based on the latter duality between Voronoi diagrams and Delaunay triangulations. Ledoux (2007) explains how to transform a valid 3D Delaunay triangulation into a valid Voronoi diagram.

### 3.2.3 *Construction of the Delaunay triangulation*

Various algorithms exist to generate the Delaunay triangulation of a given set of points. One straightforward but computationally inefficient approach is the brute-force method, which involves examining all possible triangulations. This approach costs $\mathcal{O}\left(n^4\right)$, since there are $\mathcal{O}\left(n^3\right)$ triangulations possible.

However, in the 2D case, an optimal construction algorithm operates in $\mathcal{O}(n \log n)$.

The *Flip Algorithm* in the 2D case initiates with an arbitrary triangulation and iteratively flips edges (i.e. replaces an edge with a new one) according to pre-defined rules until a valid Delaunay triangulation is achieved. This method can be extended to the 3D case, although the flipping process becomes more complex.

The *Incremental Algorithm* follows the fundamental principles of the Flip Algorithm. However, rather than starting with the entire set of points, points are added incrementally, with edge flips performed as necessary to maintain the validity of the Delaunay triangulation (hence the term 'incremental'). This approach can also be extended to three dimensions.

In the 3D case, incremental and flip algorithms are used due to their simplicity. The complexity of the incremental algorithm is quadratic ($\mathcal{O}\left(n^2\right)$). However, insertion in a certain order reduces the complexity to $\mathcal{O}(n \log n)$ in expectation (Edelsbrunner & Shah 1996). The incremental algorithm is also the implementation that is adopted in our code.

Other notable methods for Delaunay triangulation include the DeWall algorithm (Cignoni, Montani & Scopigno 1998), divide-and-conquer techniques, and a projection-based approach. The projection algorithm works by embedding the points into a higher dimensional space (by adding an extra coordinate), computing the convex hull in that space, and then projecting the points back into the original dimension by removing the added coordinate.

A comprehensive comparison of these algorithms, along with detailed implementation considerations, is provided in Elshakhs et al. (2024). This source also includes additional insights into Delaunay triangulation, its applications, and GPU-based implementations.

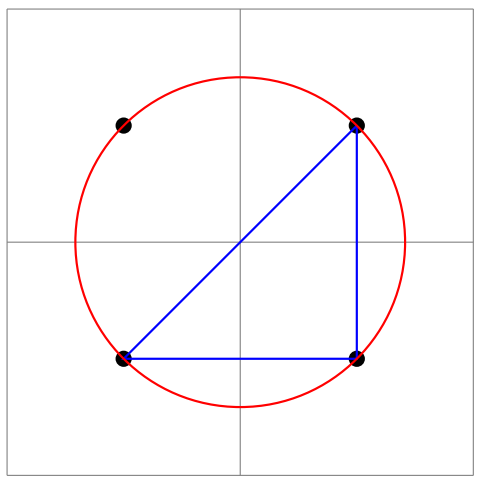

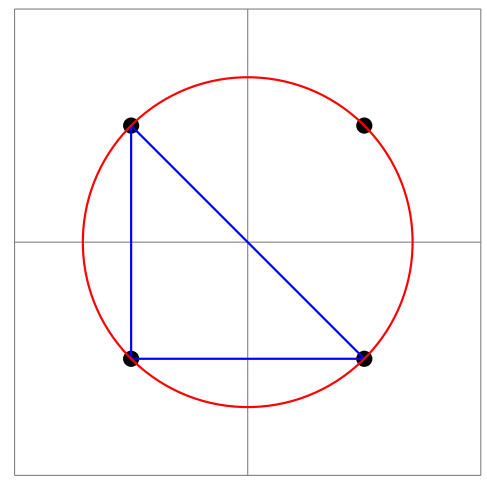

**Figure 4.** Given four points (in black), there is more than one possible Delaunay triangulation (shown to the left and the right). The two bottom points may share a triangle with both upper points. The Delaunay triangle is and its circumcircle, having a point on its boundary, are drawn.

### 3.2.4 *Mesh degeneracies*

In some cases, the circumsphere of a particular tetrahedron in 3D (or, the circumcircle of a triangle in 2D) may contain a point located on the circle itself. This situation is inevitable in some cases, for example, in a Cartesian grid, as shown in Fig. 4. As shown in this example, there are two Delaunay triangulations possible for the presented 2D Carteisan grid. We define the Delaunay triangulation to be valid even when there are degeneracies. Testing whether a point is inside or outside a sphere is done by calculating a determinant and considering its sign ($> 0$ means outside, and $< 0$ means inside). If the point is on the circle's boundary, the determinant is 0. The latter calculation is sensitive, and using regular floating-point primitives such as float or double might yield incorrect results due to their precision. Thus, it is important, in some cases, to calculate accurately. Therefore, when an accurate result is necessary, we use a library developed by Shewchuk (1997) that can calculate the result with an adaptive floating point precision.

## 4 SPRINGEL'S ALGORITHMS

### 4.1 The unified circle algorithm

Let us introduce Springel's first algorithm presented in his code AREPO (Springel 2010), a 3D hydrodynamic simulation. In this framework, the Voronoi diagram serves as the spatial discretization method, partitioning space into cells, each of which holds physical data. As the generating points move with each time-step of the simulation, the Voronoi diagram must be reconstructed (or at least partially updated). Due to the continuous nature of the system, minor movements of the points result in only minimal changes to the Voronoi diagram. The following terms are defined for convenience:

(i) $C_R(p)$: the sphere (circle) of radius $R$ around a point $p$.

(ii) $R_T$: the radius of the circumcircle of a triangle $T$.

(iii) intersect $(C_R(p))$: the processors that intersects with the sphere $C_R(p)$.

(iv) $\alpha$: the multiplicative factor of circle inflation (a hyperparameter which in our implementation is determined by default as 1.1).

**Algorithm 1** Springel's Algorithm of Building a Distributed Voronoi Diagram

**Require:** $radiuses$ array. ▷ Will be discussed later.
1: Compute a triangle containing all $points$ and locally build the Delaunay triangulation of $points$.
2: $cur_points \leftarrow points$ ▷ Remaining points to build
3: $cur_radiuses \leftarrow radiuses$ ▷ Remaining points' radiuses
4: $all_points \leftarrow points$
5: **for** $P = 0, ..., N_{proc} - 1$ **do**
6: $sent\,[P] \leftarrow \emptyset$
7: **end for**
8: **while** There's a processor that insists to continue **do**
9: $new_points \leftarrow \emptyset$
10: $reuqests \leftarrow \emptyset$
11: **for** $p$ in $cur_points$ **do**
12: $r \leftarrow cur_radiuses\,[p]$
13: **for** $P$ in $intersect(C_r\,(p))$ which is not self **do**
14: add $(P, p, r)$ to $requests$ (outcoming)
15: **end for**
16: **end for**
17: **Exchange** $requests$
18: **for** $(P, p, r)$ in $requests$ (incoming) **do**
19: $A \leftarrow (points \cap C_r\,(p)) \setminus sent\,[P]$
20: $sent\,[P] \leftarrow sent\,[P] \cup A$
21: **Send** $A$ to $P$
22: **end for**
23: **while** There's an incoming message $A$ from $P$ **do**
24: **Receive** $A$ from $P$
25: $new_points \leftarrow new_points \cup A$
26: **end while**
27: Add $new_points$ to the current Delaunay triangulation.
28: **for** $p$ in $cur_points$ **do**
29: $r \leftarrow cur_radiuses\,[p]$
30: $T_p \leftarrow$ triangles that $p$ is a part of
31: $R \leftarrow \max_{T \in T_p}(R_T)$
32: **if** $r \geq 2 \cdot R$ **then** ▷ Done
33: remove $p$ from $cur_points$
34: $radiuses\,[p] \leftarrow r$
35: **else**
36: $cur_radiuses\,[p] \leftarrow cur_radiuses\,[p] \cdot \alpha$
37: **end if**
38: **end for**
39: Call to halt if $cur_points = \emptyset$, otherwise insist to continue.
40: **end while**
41: Find the dual Voronoi diagram of the resulting Delaunay triangulation.

It is important to notice that the algorithm itself is not optimized. For example, one may consider aggregating requests in a buffer to decrease communication congestion or maintaining special data structures to compute intersections quickly.

Assuming one is persuaded that a point is removed from the $cur_points$ list only when all of its correct Delaunay neighbours arrive at the processor, the correctness is evident. This property is correct since we remove the point only when its test circle encompasses all of its circumcircles (line 33). At this point, we have already brought all the points within the test circle and, consequently, all points inside the circumcircles as well. Therefore, the circumcircles are free of any foreign points that have not yet been brought in. To facilitate the successive constructions, Springel maintains a list of the final search radii from Algorithm 1. This list is then used as an initial estimate for each point's search radius in subsequent constructions, reducing the number of required iterations. The difference in running time is noticeable. Consequentially, the first construction typically takes considerably longer than subsequent builds. In this context, the key metric is the construction time for the advanced builds, as they account for the majority of the simulation's total running time.

#### *4.1.1 Demonstration*

We illustrate the steps of Springel's algorithm through a simple example. For clarity, we'll have the example in the 2D space (although our algorithm operates in three dimensions). In addition, we will deal with two processors only (the red at the top and the blue at the bottom) and focus on a single point. In an original single iteration, the algorithm runs for all local points (unless, of course, they are done) and, in particular, for our point.

Consider Fig. 5(a), which depicts the first iteration of the build process. An initial circle is placed around the local point (the red circle). In this phase, the circle neither intersects the blue processor nor contains any local points other than the centre point. Consequently, no new points are brought in. Additionally, we have not completed the build process for our point because its Delaunay circles (shown in grey) are not fully contained within the red circle. Therefore, we proceed to the next build iteration, in which the testing circle is increased by $\alpha$.

Next, we move to the situation illustrated in Fig. 5(b). The previous phase circle is drawn in dashed purple. We now ask for points located inside the red circle. This time, there is a local point matching the query (located right below our point), but since it's a local point and already a part of our local Delaunay triangulation, there is no need to do something. Again, no new points are found, and we cannot finish the build process for the point, as the red circle does not contain all the Delaunay circles (in grey). Therefore, we enlarge the circle once more. Refer to Fig. 5(c). We scrutinize the points inside the red circle. This time, the circle intersects with the blue processor, prompting us to send a remote range query. In response, the remote process sends us a new ghost point. We build a new Delaunay triangulation (in fact, we modify the existing one) to incorporate this newly received ghost point. Since the Delaunay circles before the latest build (in grey) are still not included in the red circle, we inflate it even more. In Fig. 5(d), we reiterate this procedure, retrieving two additional remote points. We construct a new Delaunay triangulation and increase the red circle radius, as the old Delaunay circles are not contained in the red one. Turning to Fig. 5(e), all requisite ghost points have already been assimilated. Nevertheless, the termination criterion is not yet fulfilled. Two additional points are incorporated, albeit superfluously. As we can see in Fig. 5(f), the iterations for our point are done, as the current circle contains all the grey circumcircles (the point triangle's circumcircles). We can remove our point from the list of unfinished points (*cur_points* in algorithm 1).

### 4.2 Individual circles approach

Recall that the algorithm asks for all the ghost points inside the big testing circle, imports them, and then uses all to build a Delaunay triangulation. However, creating excessive layers of unnecessary ghost points, which might occur in multiple cases (as we will demonstrate soon), may severely degrade efficiency. Springel (2010) proposed an alternative approach, where instead of testing a big circle, attempting to capture at once all the points inside this point

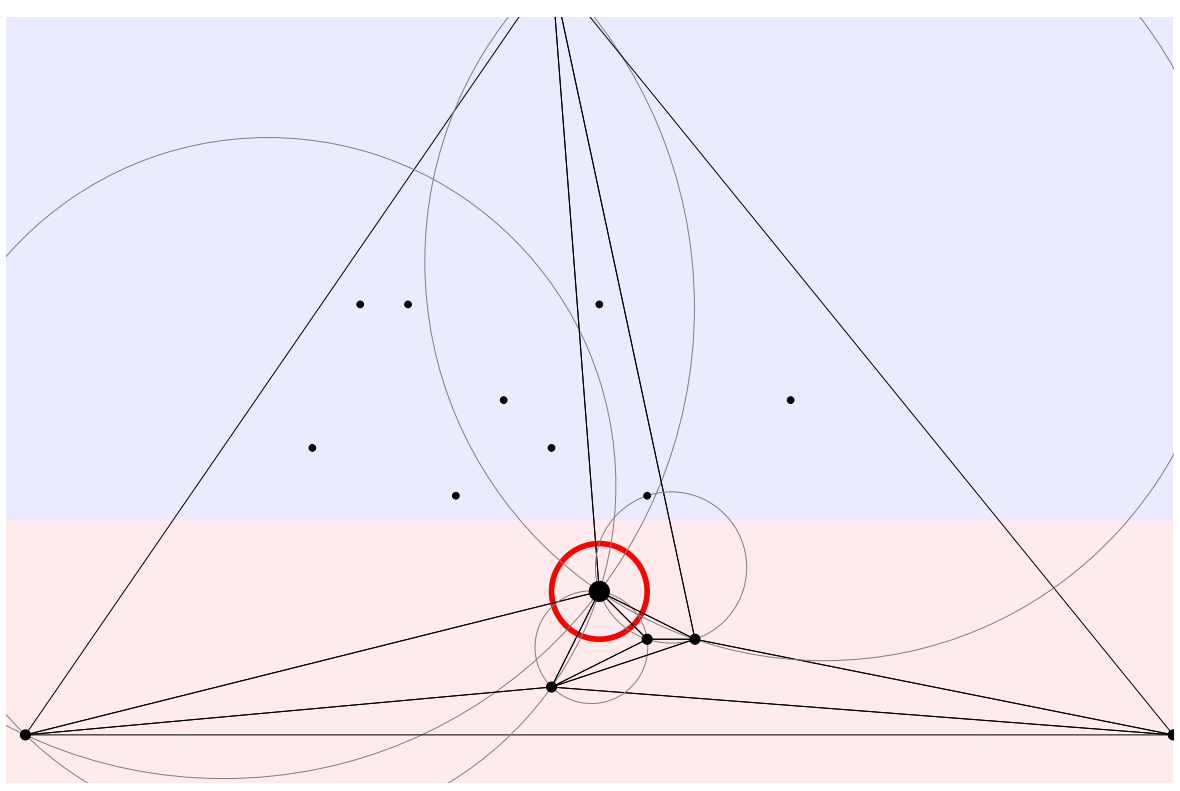

(a) An initial circle is drawn around the point. Current circumcircles (in gray) are not contained, so we increase it.

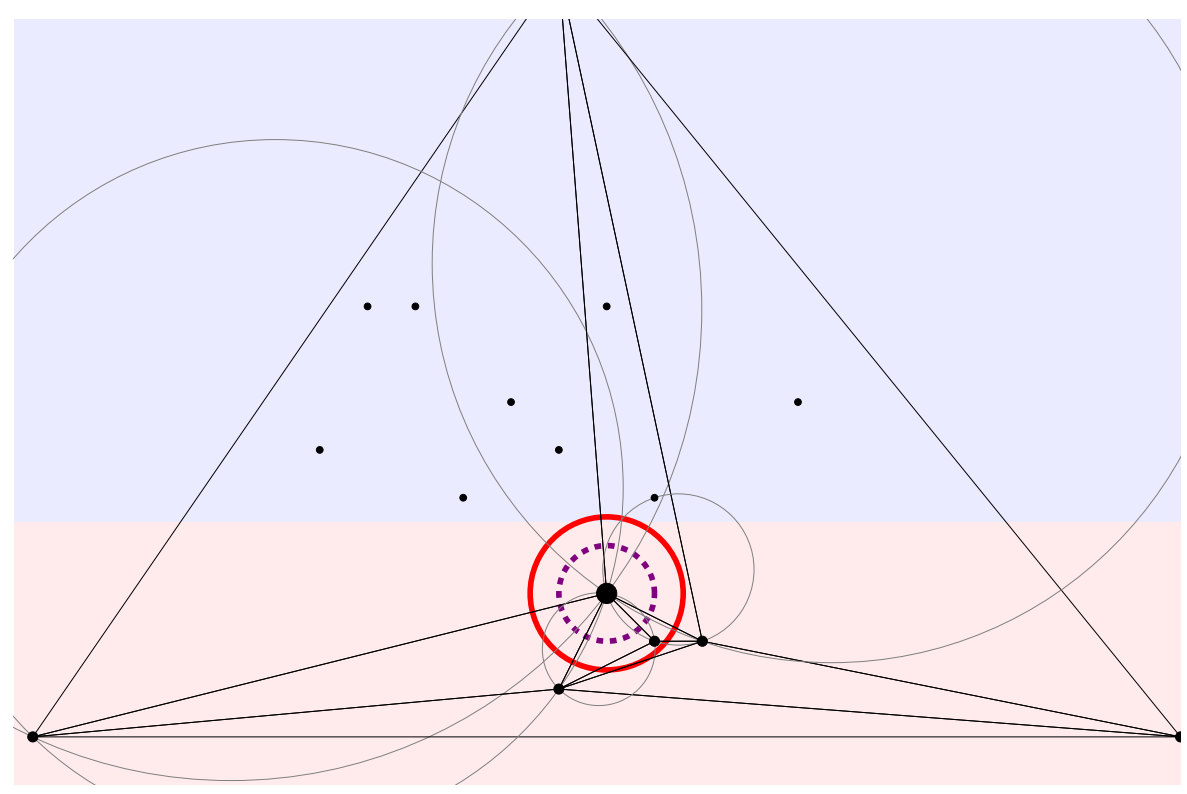

(b) We bring all the points inside the blue area and the red circle, and then we build the Delaunay triangulation again. However, no points were brought in this step.

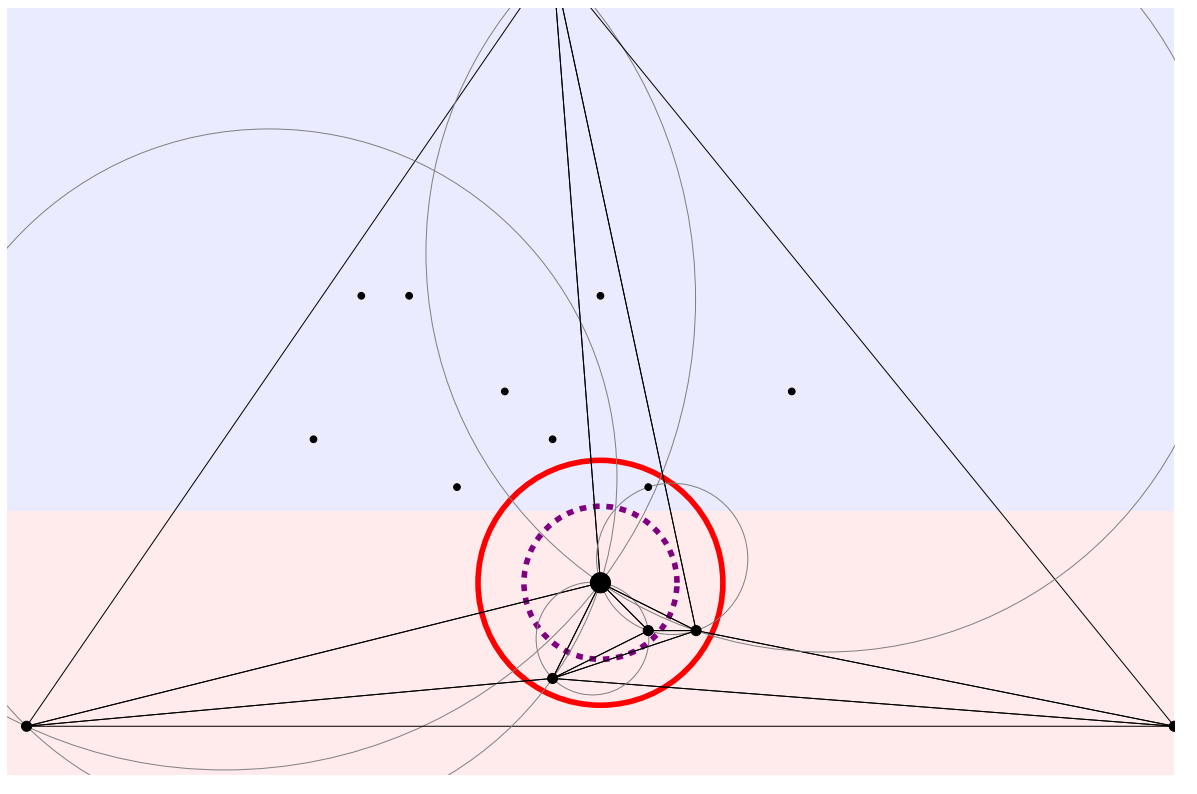

(c) We increase the red circle's radius since it didn't contain all the gray circles. One more remote point is brought and added to the Delaunay triangulation.

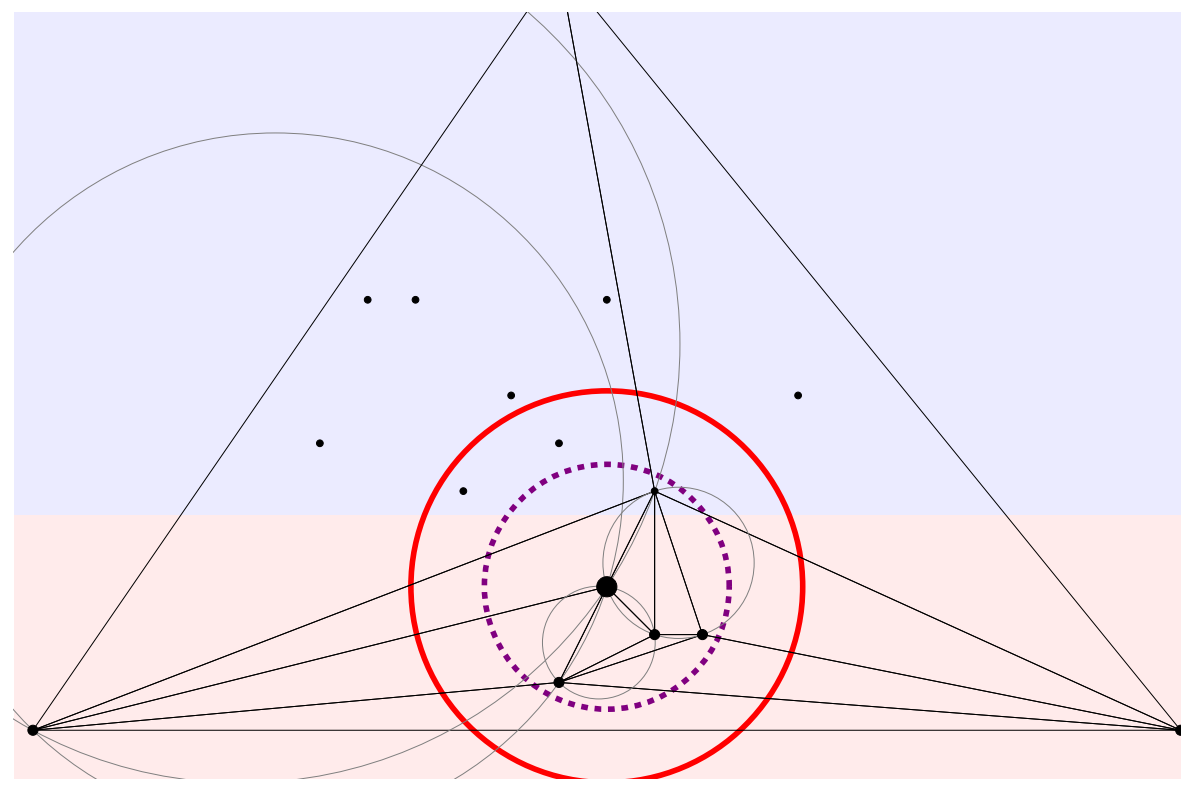

(d) We increase the red circle again. Two more points are added to the triangulation.

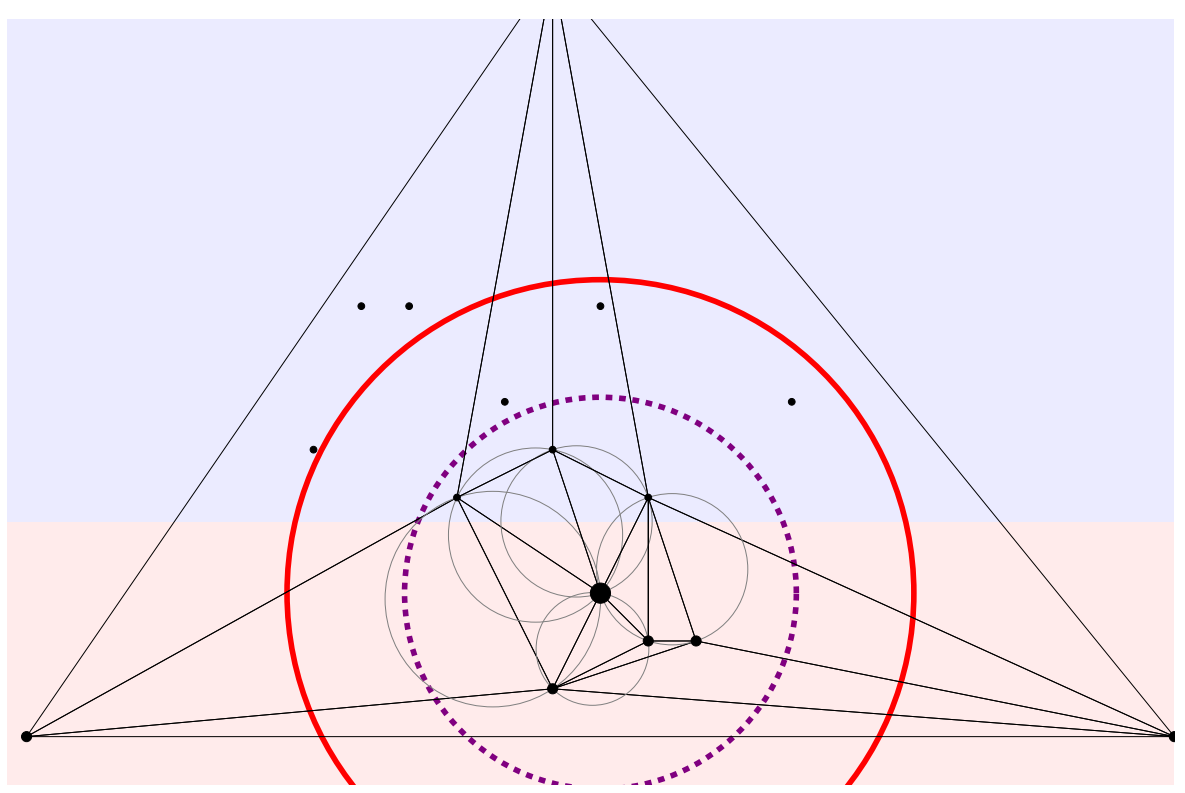

(e) Even though the current point triangles are legal, the gray circles are not all contained in the red circle. So we'll increase it one last time. Three more points are added. Notice that they are all redundant points.

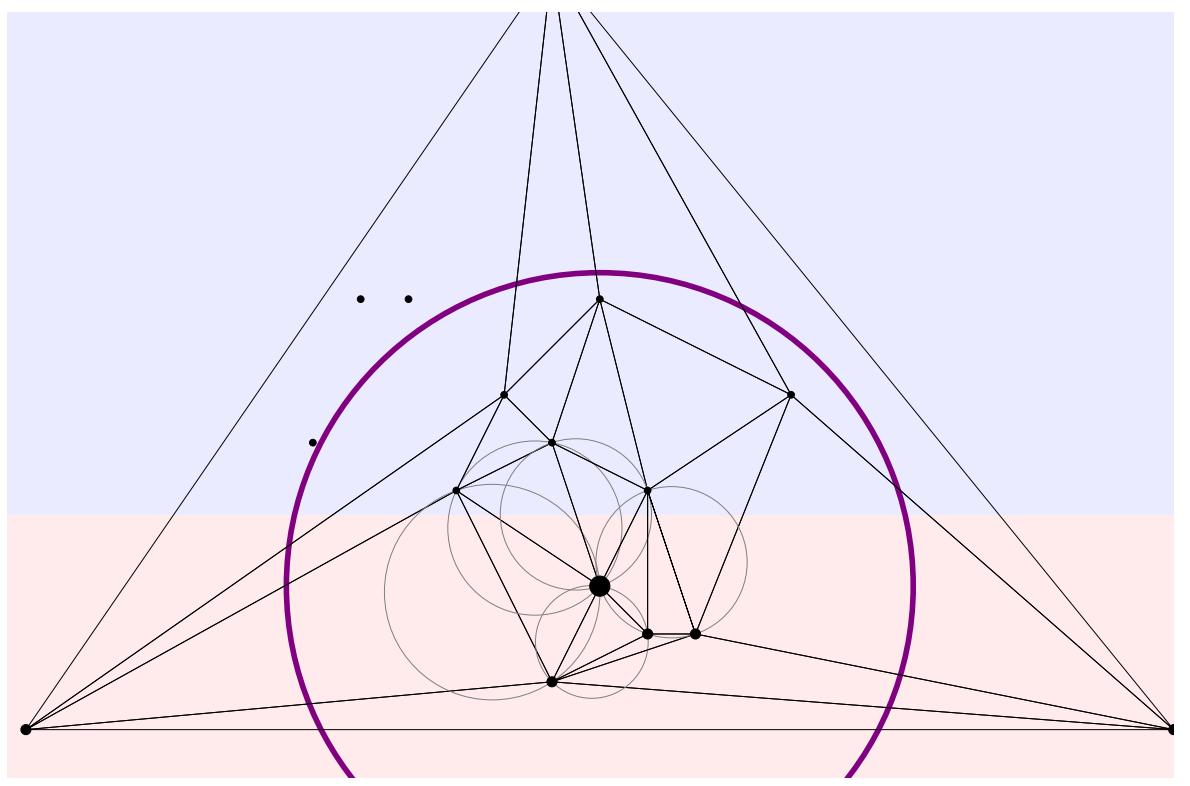

(f) Since the red circle contains all the gray circumcircles, we are done for this point.

**Figure 5.** Example for running algorithm 1 for one point only and two processors (red and blue).

triangles' circumcircles, we instead break the range query into multiple alternative queries, each corresponds to one circumcircle only. In other words, we ask the relevant processors to bring all the points inside each circumcircle separately. The algorithm is described in 2. Notations are the same as in algorithm 1.

In algorithm 2, we iterate over any point, say $p$, then over all of $p$'s circumcircles and ask the intersecting processors for their closest point to $p$ inside the circumcircle. We add imported points to the Delaunay triangulation. Instead of retrieving an unbounded number of points in a big circle, we examine the circumcircles separately.

The advantage of algorithm 2 is that it prevents a situation in which the circles around the points are inflated unnecessarily. This situation may happen if the triangles of a point in a legal Delaunay triangulation are not uniform in size, as shown in Fig. 6.

**Algorithm 2** Alternative Algorithm for Distributed Construction of Voronoi Diagrams

1: Compute a triangle containing all $points$ and locally build the Delaunay triangulation of $points$.
2: $cur_points \leftarrow points$
3: $cur_triangles \leftarrow \bigcup_{p \in points} T_p$ ▷ All triangles
4: $new_triangles \leftarrow cur_triangles$
5: $all_points \leftarrow points$
6: **for** $P = 0, ..., N-1$ **do**
7: $\quad sent\left[P\right] \leftarrow \emptyset$
8: **end for**
9: **while** There's a processor that insists on continuing **do**
10: $\quad reuqests \leftarrow \emptyset$
11: $\quad triangles_checked \leftarrow \emptyset$
12: $\quad$ **for** $p$ in $cur_points$ **do**
13: $\quad\quad$ **for** $T$ in $T_p \cap cur_triangles$ **do**
14: $\quad\quad\quad r \leftarrow R_T$
15: $\quad\quad\quad c \leftarrow$ the center of the circumcircle of $T$
16: $\quad\quad\quad$ add $(P, p, c, r)$ to $reuqests$ (outcoming)
17: $\quad\quad$ **end for**
18: $\quad$ **end for**
19: $\quad$ **Exchange** $requests$
20: $\quad$ **for** $(P, p, c, r)$ in $requests$ (incoming) **do**
21: $\quad\quad A \leftarrow C_r(c) \setminus sent\left[P\right]$
22: $\quad\quad p' \leftarrow argminx \in Ad(x, p)$
23: $\quad\quad sent\left[P\right] \leftarrow sent\left[P\right] \cup \left\{p'\right\}$
24: $\quad\quad$ **Send** $p'$ to $P$
25: $\quad$ **end for**
26: $\quad$ **while** There's an incoming message $p'$ from $P$ **do**
27: $\quad\quad$ **Receive** $p'$ from $P$
28: $\quad\quad all_points \leftarrow all_points \cup \left\{p'\right\}$
29: $\quad$ **end while**
30: $\quad$ Build a Delaunay triangulation of $all_points$ and mark new triangles
31: $\quad new_triangles \leftarrow \emptyset$
32: $\quad$ **for** $p$ in $points$ **do**
33: $\quad\quad$ **for** $T$ in $T_p$ **do**
34: $\quad\quad\quad$ **if** $T$ is new **then**
35: $\quad\quad\quad\quad new_triangles \leftarrow new_triangles \cup \{T\}$
36: $\quad\quad\quad$ **end if**
37: $\quad\quad$ **end for**
38: $\quad$ **end for**
39: $\quad$ Call to halt if $new_triangles = \emptyset$, otherwise insist to continue.
40: **end while**
41: Find the dual Voronoi diagram of the resulting Delaunay triangulation.

## 5 LOAD BALANCING

To build the Voronoi diagram in parallel, it is imperative to consider changing the distribution of generating points across hosts. This way, we allocate a subspace (or multiple subspaces) for each host. The host is responsible for exclusively constructing the Voronoi cells of the points inside its designated regions.

Better distribution diminishes communication overhead for several reasons. First, for many points, all neighbours might be assigned to the same host, and no further communication is needed in the algorithm as the cell can be built locally. Secondly, a well-balanced distribution minimizes interhost dependencies, as fewer points will have neighbours spanning multiple hosts. Thirdly, the nodes' physical topology and rank assignment should be considered; if two nodes can communicate efficiently, their assigned subspaces should also be spatially close to reduce communication costs. In MPI, ranks close to each other are often assigned (by the MPI implementation) to physically proximate hosts, improving communication speed. Therefore, we seek continuity in zone assignments, so spatially close zones will be assigned to processors with close ranks.

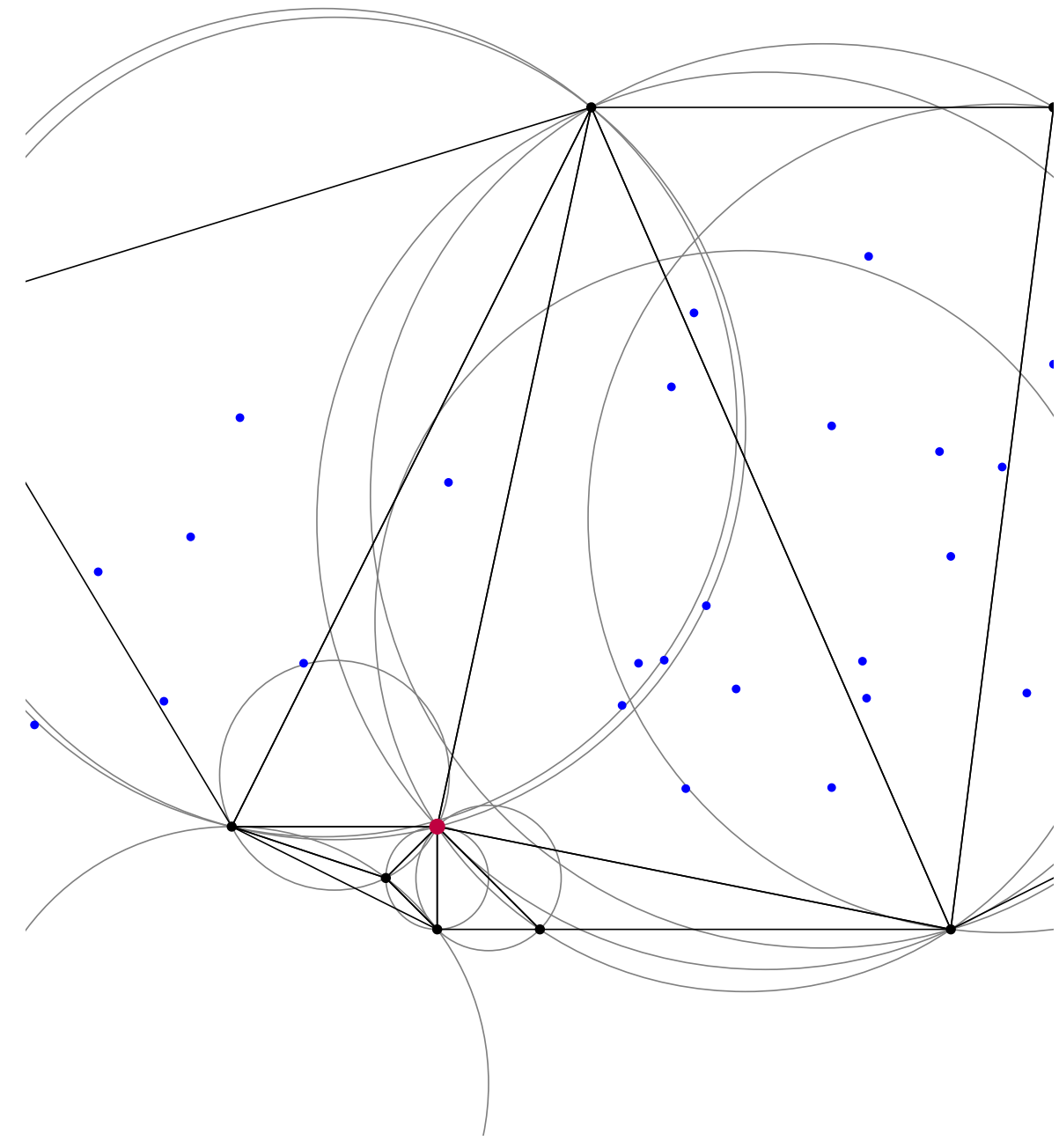

**Figure 6.** The circles of the purple point in the current local Delaunay triangulation have a large variance. Applying Springel's algorithm might bring all the remote points (blue points) into one of the big circles. A new Delaunay triangulation will be built using all the new points. However, only a small fraction of the blue points are essential.

The mesh partitioning problem, assuming one seeks to minimize communication, is equivalent to the graph partitioning problem, where the goal is to partition a given graph's vertices into disjoint sets, minimizing the number of edges connecting nodes of different sets (these are called the cut edges). Unfortunately, this problem and its similar variations are NP-hard, rendering the pursuit of efficient exact solutions impractical. Consequently, a variety of algorithms and heuristics have been developed, which are briefly outlined in Section 2.1.

We implement a heuristic called curve-based load balancing.

### 5.1 Curve-based load balancing

Before describing the curve-based load balancing, let us recall that in Voronoi diagrams, we can refer to a generating point as an equivalent to the cell it induces and vice versa. So, it is sufficient to distribute the points to the processors efficiently.

We begin by assuming an initial poor distribution of points across processors (for example, each participant maintains a random partial list of the points in space) as shown in Fig. 7(a).

In curve-based load balancing, we define a curve that traverses multiple pre-defined points (which are typically distinct from the Voronoi generating points), which will be called *curve points*. Each curve point is assigned an order based on its position along the curve. Each generating point is then mapped to its closest curve point and inherits its corresponding order.

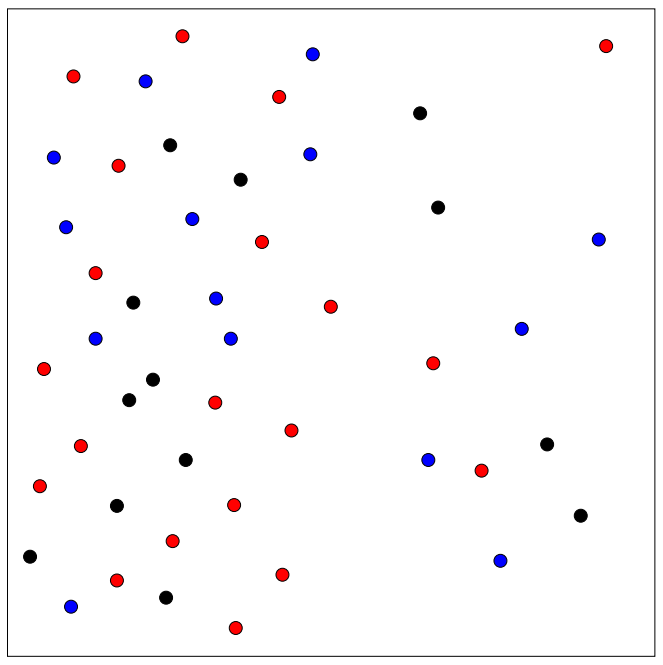

(a) Assuming each processor holds a subset of the points.

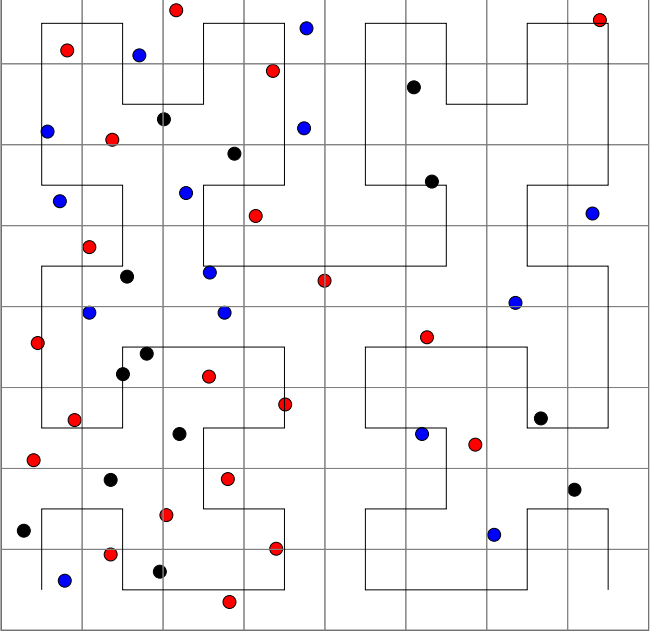

(b) We divide into square subspaces and create a (two dimensional) Hilbert curve passing through them.

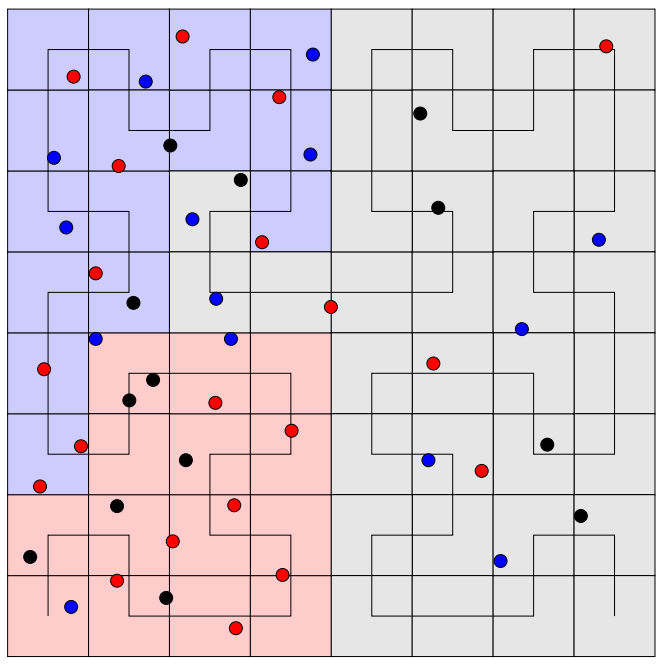

(c) Step 1 - Determine the curve cut points (balance).

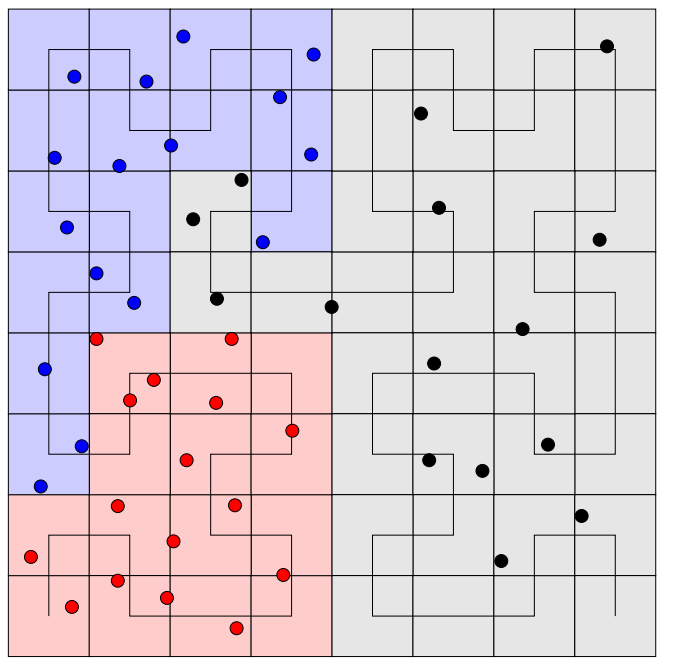

(d) Step 2 - Exchange points according to the partitioning.

**Figure 7.** Example for the load balancing process in case of 48 points and 3 processors (red, blue, green). The colour of the point expresses its ownership. The ownership of the curve cells – where all points within a cell are allocated to a specific processor – is indicated by the background colour.

Each processor maintains a local list of numbers representing the enumeration of the closest curve point for each of its initial generating points. Ideally, these local lists would be combined into a global list containing enumeration numbers for all points across processors. If feasible, this global list is then locally sorted and divided into $N$ equally sized parts, which can be viewed as numerical ranges due to the sorting.

As illustrated in Fig. 7(c) where, after this process, domains are allocated to processors according to the curve. Then, as shown in Fig. 7(d), each rank sends each one of its initial generating points to the processor responsible for its assigned numerical range.

The quality of space-filling curves and the metric used for evaluation is contingent upon the specific problem at hand. However, space-filling curves are widely used due to their *locality preserving properties*: Close points tend to have close enumerations. Unfortunately, assembling the entire list of numbers into a single global array on one processor and performing local sorting and division is infeasible due to memory limitations. The challenge remains to determine the partition boundaries of this global array.

It is important to mention that there are techniques for distributed memory sorts as well. Yet, re-evaluating the problem, it becomes clear that a full sort is unnecessary. Assuming there are $P$ processors and $n$ numbers in total, we merely wish to find the smallest values at positions $n/P$, $2 \cdot n/P$, ..., $P \cdot n/P$ in the global array to define the partition boundaries.

The $k$-smallest value in an array is called the *kth-order statistic* of the array, and the task of finding it is called a *selection* problem. We developed a technique for efficiently finding those statistics in a distributed memory setting.

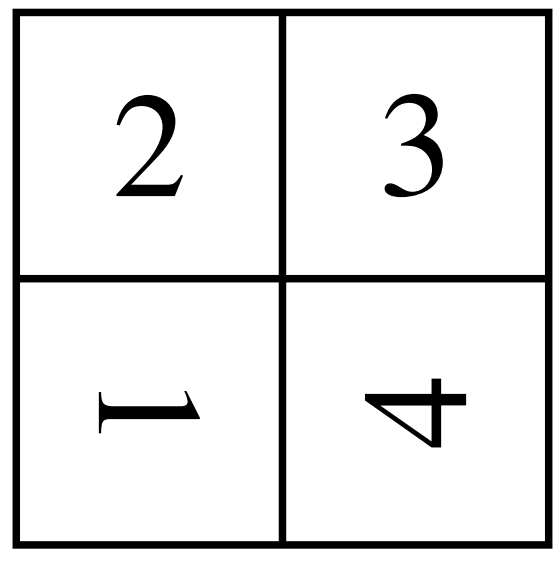

**Figure 8.** A division of the unit square into four subsquares.

Our suggested order statistics finding algorithm generalizes the known *QuickSelect* algorithm to distributed memory systems. The QuickSelect algorithm is known to find the $k$-th order statistic of an array by recursively partitioning the array until the desired element is found. The algorithm can be generalized into a distributed memory system by using synchronization and collective communication operations such as *Allreduce* to determine the current pivot element order statistic. We also generalized the algorithm to find all the desired statistics (these are, $n/P$, ..., $P \cdot n/P$) in a single run.

This method allows us to find the exact partition points of the curve.

We further extended the algorithm to accommodate weighted partitions. In case one does not want to distribute the points equally, a weight function can be assigned to the points, and the algorithm then computes a partition in which the total weight assigned to each processor – calculated as the sum of the weights of the points allocated to it – remains approximately balanced.

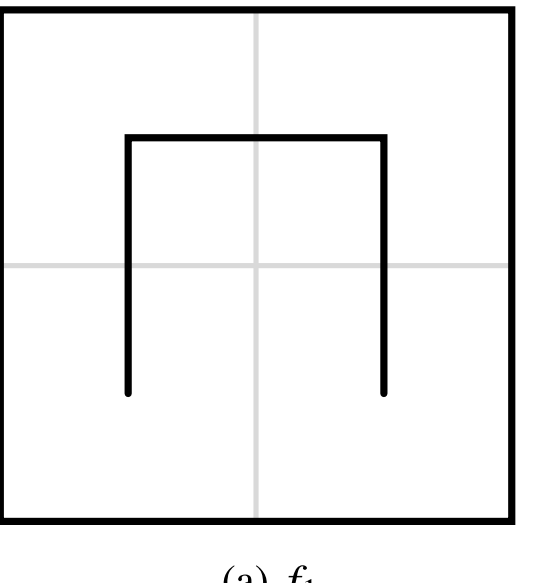

(a) $f_1$

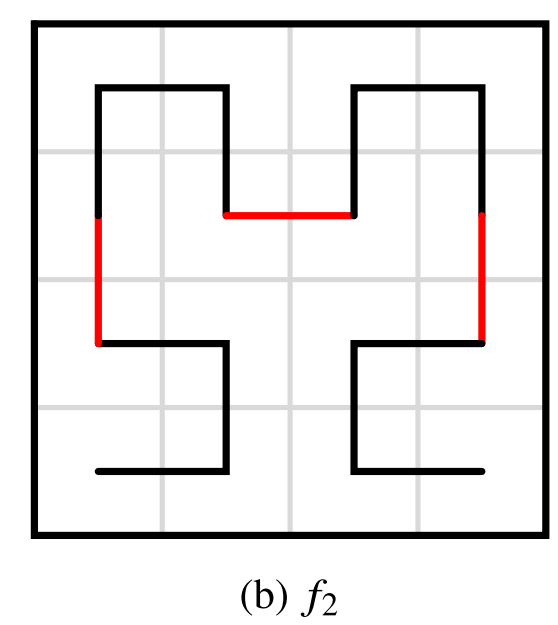

(b) $f_2$

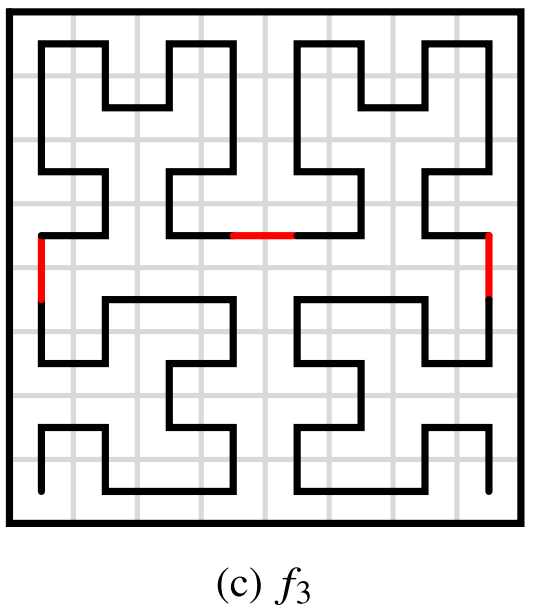

(c) $f_3$

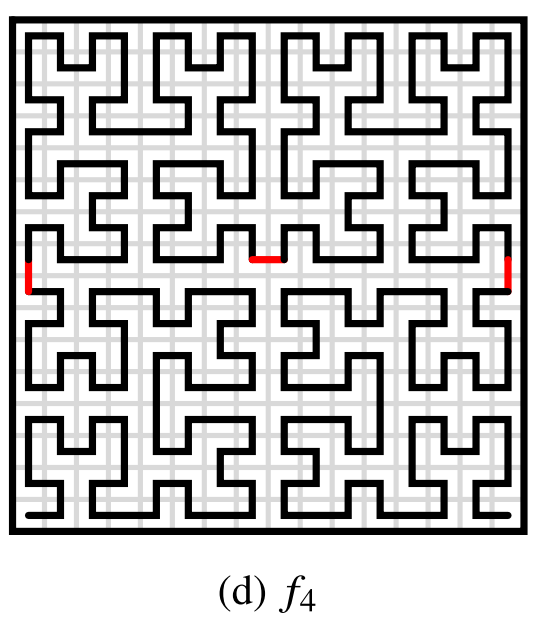

(d) $f_4$

**Figure 9.** Example for the first Hilbert curves in the sequence. Recursive division lines (to subsquares) are shown in grey.

### 5.2 Hilbert curve

Our code uses curve-based load balancing where the enumeration is based on the Hilbert curve. This curve is recursively defined. As an example, let us define the 2D Hilbert curve. A *Hilbert curve* is a sequence of curves $f_0, f_1, \ldots$, recursively defined as follows:

(i) $f_0 \equiv \left(\frac{1}{2}, \frac{1}{2}\right)$.

(ii) Let $n \geq 0$, and assume that $f_n$ was defined. $f_{n+1}$ (also called *the* $(n+1)$*-level*) is defined as the following: split the space $[0, 1]^2$ into 4 squares, as instructed in Fig. 8. Hold, for each subsquare, a copy of $f_n$. Rotate the copy of subsquare 1 $90^\circ$ clockwise, and rotate subsquare 4 $90^\circ$ anticlockwise. Then, connect each subsquare's endpoint to the next one's beginning.

The function $f_n$ is called a *Hilbert curve of order n*. Examples for $f_1, f_2, f_3, f_4$ are shown in Fig. 9.

It is possible to generalize the Hilbert curve to higher dimensions, where the 3D curve is often referred to as Peano-Hilbert curve. The main advantage of using the Peano-Hilbert curve over other curves is that this curve preserves locality effectively. Informally, that is, given an order $n$, the curve $f_n$ ensures that if $x \approx y$, then $f_n^{-1}(x) \approx f_n^{-1}(y)$ (for all $x, y \in \mathrm{Im}(f_n)$). This attribute is not a trivial or easily achieved property, as the curve may follow a lengthy path when transitioning between points that are spatially close.

We used curve-based load balancing, using the Peano-Hilbert curve in our implementation. Our implementation uses a generalized curve, which is similar to the Peano-Hilbert curve, but only for rectangular (instead of squared) areas.[10] To calculate enumeration on the curve, we use a data structure similar to an R-tree, which is a tree maintaining the Peano-Hilbert subdomains recursively, along with each one's range of enumerations.

The load-balancing approach in a generalized rectangular space is also crucial for supporting Voronoi diagram construction within arbitrary boundaries, assuming they create a convex polygon, instead of a conventional bounding box, as load balancing becomes more challenging in irregular domains. The boundaries are defined by their vertices, and the Voronoi diagram is computed accordingly. For example, Fig. 10 illustrates a Voronoi diagram constructed within a pyramidal domain. It is important to emphasize that the computational domain itself is shaped like a pyramid, rather than just the point distribution, making the construction significantly more complex.

[10] https://github.com/jakubcerveny/gilbert

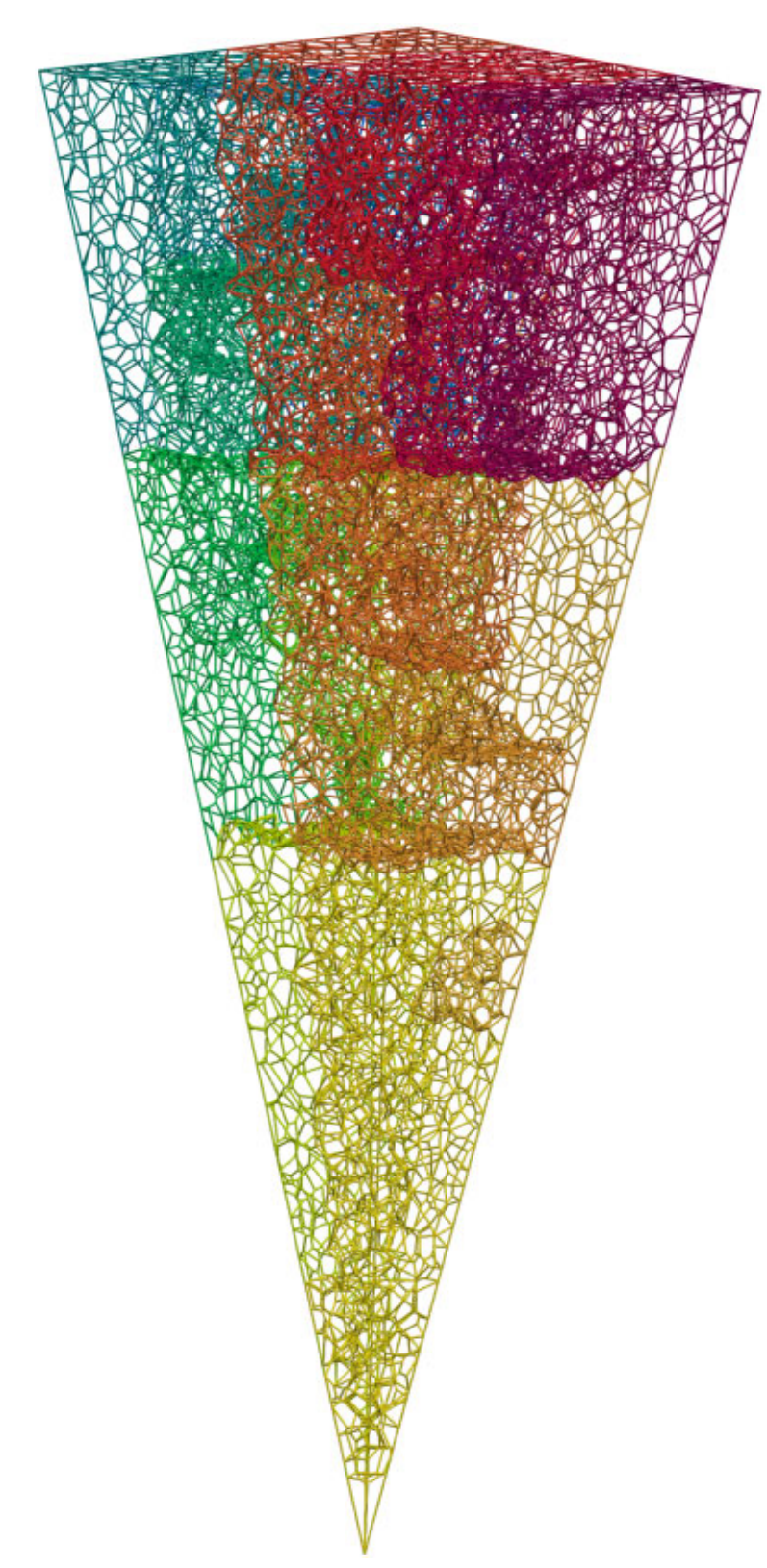

**Figure 10.** An example for the construction of a pyramid-shaped space. Voronoi cells are clipped by their corresponding boundary face, if needed.

To facilitate load balancing, we enclose the entire computational space within a bounding box, which is necessary for computing the generalized Peano-Hilbert curve. However, this bounding box is only used for partitioning and does not influence the actual Voronoi construction. Additionally, the technique of *kernelization*, introduced in Mizrachi et al. (2024), may provide a more efficient load-balancing strategy for handling complex geometrical domains.

## 6 INTRODUCING MADVORO

Both Springel's presented algorithms 1 and 2 are capable of constructing a Voronoi diagram. Our proposed method will be to integrate the methods dynamically. To do so, we categorize the points to *small points* and *large points*. A point is classified as *small* as long as the testing circle around the point does not bring a lot of unnecessary ghost points. The moment too many points are brought, it will be considered as *large*. The terminology does not refer to the actual sizes of the circumcircles, but rather to the spatial density of the region surrounding the point. A large point likely corresponds to circumcircles with unequal radii, as the testing radius has incorporated too many points, often due to issues such as those illustrated in Fig. 6. Just like both algorithm 1 and algorithm 2, our construction process follows an iterative approach, where additional points are incrementally introduced and incorporated into the Delaunay triangulation. However, in this method, the specific algorithm applied to each point depends on its classification. For small points, we employ the unified circle approach as described in algorithm 1. Once a point is classified as large, we alternate to the technique used in algorithm 2, where each one of the point's circumcircle is checked individually for ghost points. We call the points for small points *the small algorithm* and for large points *the large algorithm*.

To enhance performance, we introduced several optimizations to the algorithm. First, because a circle might grow substantially and return an overwhelming number of points, we bound the small algorithm queries to return, at most, a constant maximum number (say, 15) for each query. If a query exceeds this threshold, the point is reclassified as large, thereby switching the algorithm used in subsequent iterations.

Secondly, when running large points queries, difficulties may arise if the triangle's circumcircle is excessively large. In such cases, the circumcircle may intersect multiple processor domains, requiring queries to each of them. In small queries, that is less likely to happen because as soon as too many points are brought, the point changes its classification to large. Instead of immediately querying all processors intersected by a large query, we adopt a two-phase approach. Initially, we query only the nearest intersecting processors. In a subsequent iteration, we extend the request to all intersecting processors. A large query is considered complete only after passing both phases. In many cases, querying only the nearest processors yields sufficient points to construct the Delaunay triangulation in the same iteration, and that makes the circles for the next iteration small. In other words, the large circumcircle may be eliminated in the process, potentially making the second check less expensive. Like Springel's first algorithm (algorithm 1), we store the final search radii at the end of the construction to facilitate later constructions. The radius saved to points is determined by the last iteration in which the point was classified as small. If the point was moved to be classified as large, we multiply this radius by a factor smaller than 1 to prevent the circle from inflating significantly in future builds. Without this adjustment, the radius – used as the initial estimate for subsequent constructions would continuously and exponentially expand, potentially degrading performance.

### 6.1 Demonstration

We demonstrate our algorithm in Fig. 11. As in Fig. 5, we present a 2D example and apply the algorithm to a single point to facilitate comprehension.

In Fig. 11, we focus on the red processor, having the two blue and green neighbours.

Let us assume the threshold of a point status change is by bringing three or more points in a single iteration.

In the beginning, the red processor is unaware of any point outside of the red area, excluding the big triangle that bounds the whole space. It starts by building a Delaunay triangulation of the points it is aware of only. Fig. 11(a) starts by assigning the point to a *small point* status, so an initial testing circle is drawn around the point. The small status remains for several iterations, and hence, the testing circle grows exponentially, intersecting the blue and then the green processors. However, no new points are brought until a further iteration, shown in Fig. 11(b). In this iteration, two points are brought up. Those points are replicated in the red processor, and added to the current Delaunay triangulation as shown in Fig. 11(c), depicting the start of the next iteration where the red testing circle grows again since the building is not complete (the red circle does not contain all the grey circumcircles). After multiple iterations, Fig. 11(d) shows the retrieval of three additional points: one from the blue processor and two from the green one. Since the number of points brought in this iteration is 3, which is the threshold of status change, we change the point status to be *large*. This status change is reflected in Fig. 11(e), where after adding the newly acquired points into the local Delaunay triangulation, the final testing circle (depicted as a dashed purple line) is drawn. From this point onward each of the point's circumcircles is tested individually. Recall that in the large phase of a point, each processor only gives the closest point to the query point, located inside each circumcircle it intersects. A tested circumcircle will be emphasized in orange. In 11(e), all the circumcircles are tested for an accurate start, and hence they are all orange. While most of the tested circles return empty, one of them retrieves a ghost point located in the blue processor, which is imported. Again, this point is added to the local triangulation. New triangles created by this point will have their circumcircle tested in the following iteration, as shown in Fig. 11(f). At this stage, all tested circumcircles are found to be empty, confirming that no further points need to be added. Consequently, the construction is complete, as shown in Fig. 11(g).

This example highlights a key limitation of algorithm 1: its exponential circumcircle growth, which can eventually encompass the entire domain. This issue arises because the given point forms an obtuse-angled triangle with an extremely large circumcircle that must ultimately be contained within the testing circle. The given example underscores the limitations of algorithm 1 against algorithm 2 in this particular example. Yet, using algorithm 2 requires more memory and communication (since each circumcircle is tested individually). As discussed in Section 7, this approach may introduce other complications. We believe our approach successfully resolves this challenging issue by leveraging the advantages of both algorithms, thereby optimizing performance while minimizing their respective limitations.

## 7 EVALUATION

We present three types of evaluations for our framework. First, we compare the construction time using two different data sets of points, running each framework on a single shared-memory machine. Secondly, we evaluate our framework in a distributed-memory setup, comparing it to other frameworks while varying the number of points per processor, keeping the number of processors constant. The third evaluation type focuses on scaling, where we examine both weak and strong scaling of our code.

All tests were executed on Linux machines running Rocky 9.3 with kernel version 5.14.0, powered by Intel(R) Xeon(R) Gold 6434 processors (2 sockets). Each node is equipped with 64 GB (DDR 5) in main memory and features Infiniband with a throughput of 100 $\mathrm{Gb\,s^{-1}}$ (2 HDR links). Hyperthreading was disabled. We compiled all codes using the Intel Compiler (2024.2.1) and employed IntelMPI as our MPI implementation.

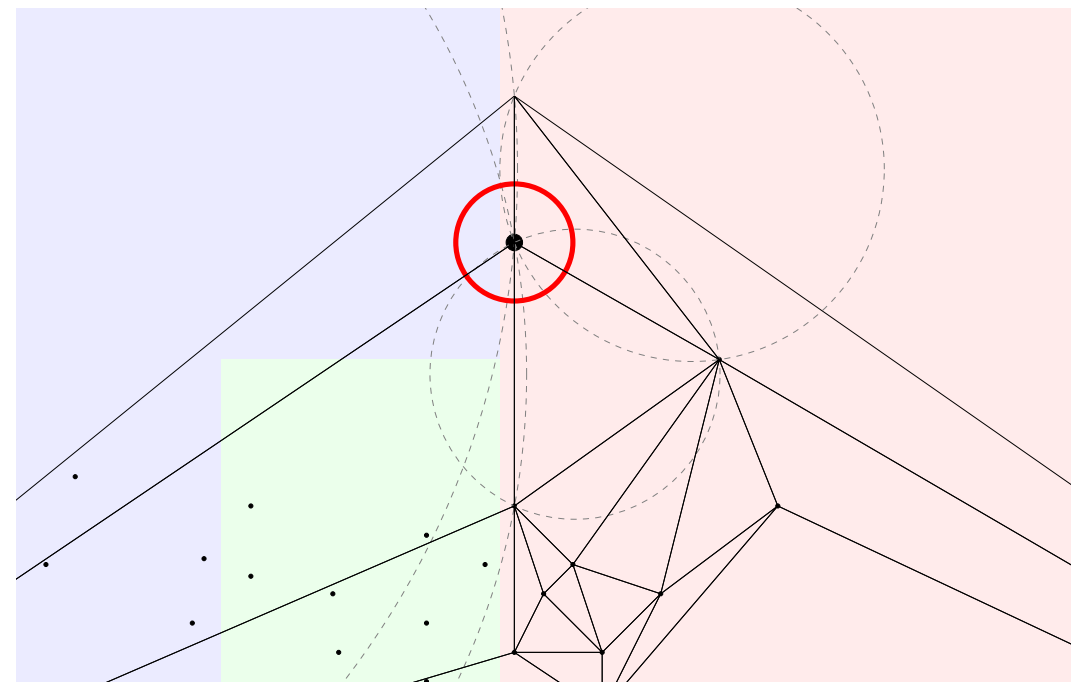

(a) The point starts by being classified as a small point. We draw a testing circle around it.

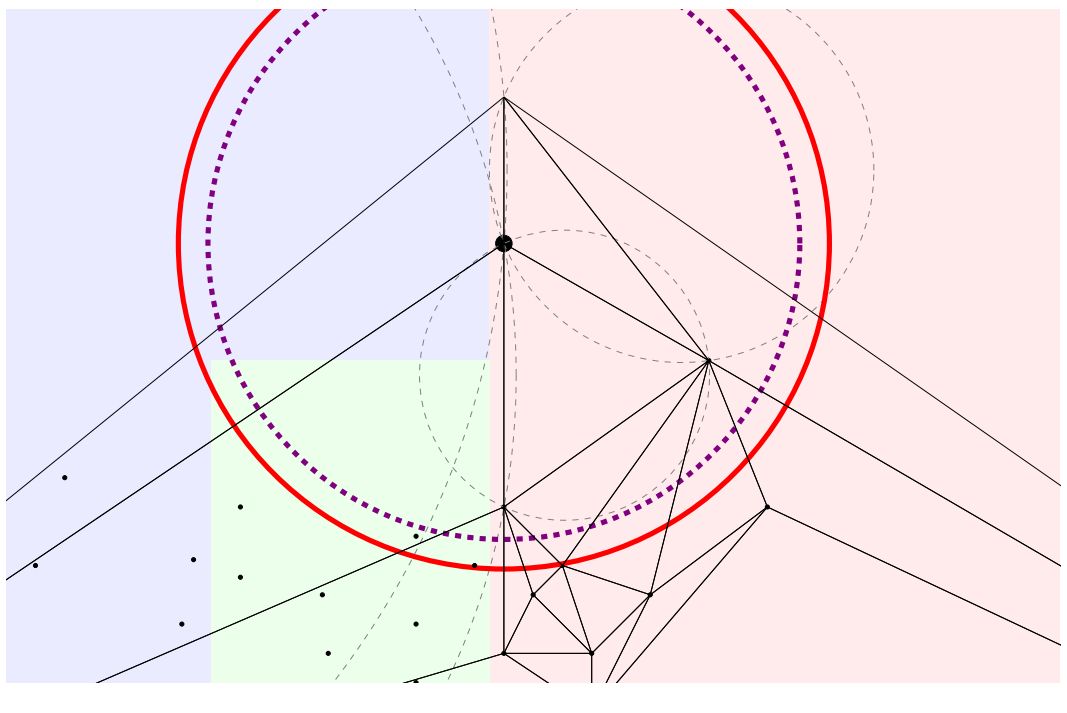

(b) In each iteration, we ask intersecting processors for all their points inside the testing circle. Only in iteration 18 two more points are brought.

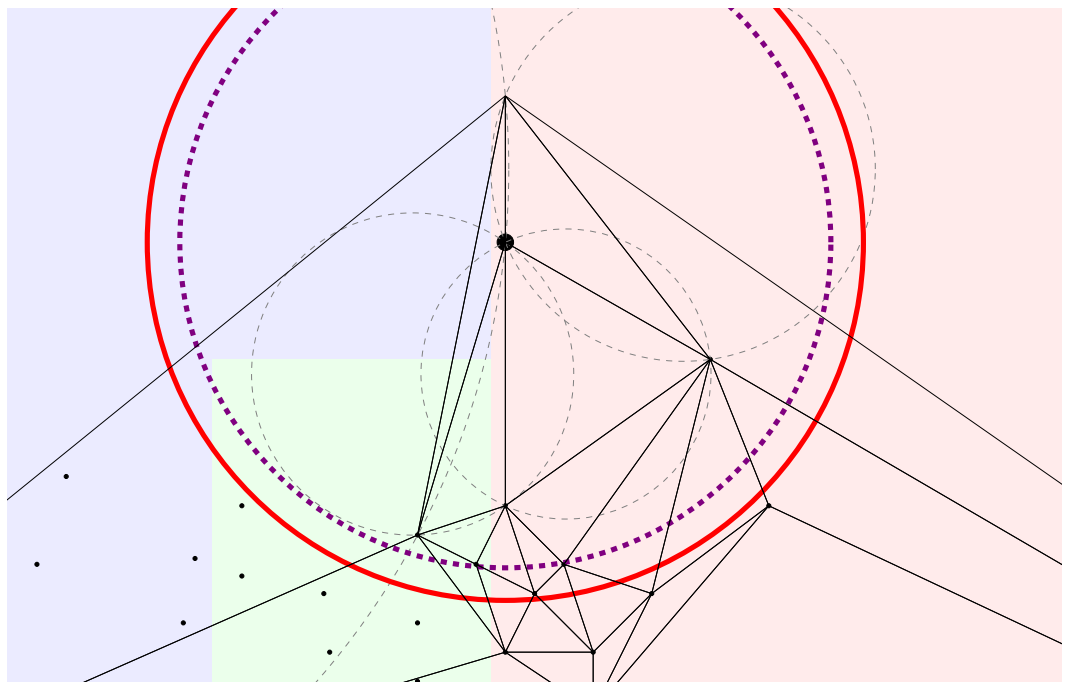

(c) The incoming points are brought and added to the local Delaunay triangulation. These are only 2 points (below of the threshold of 3). The testing circle does not contain the gray circumcircles, so we increase it.

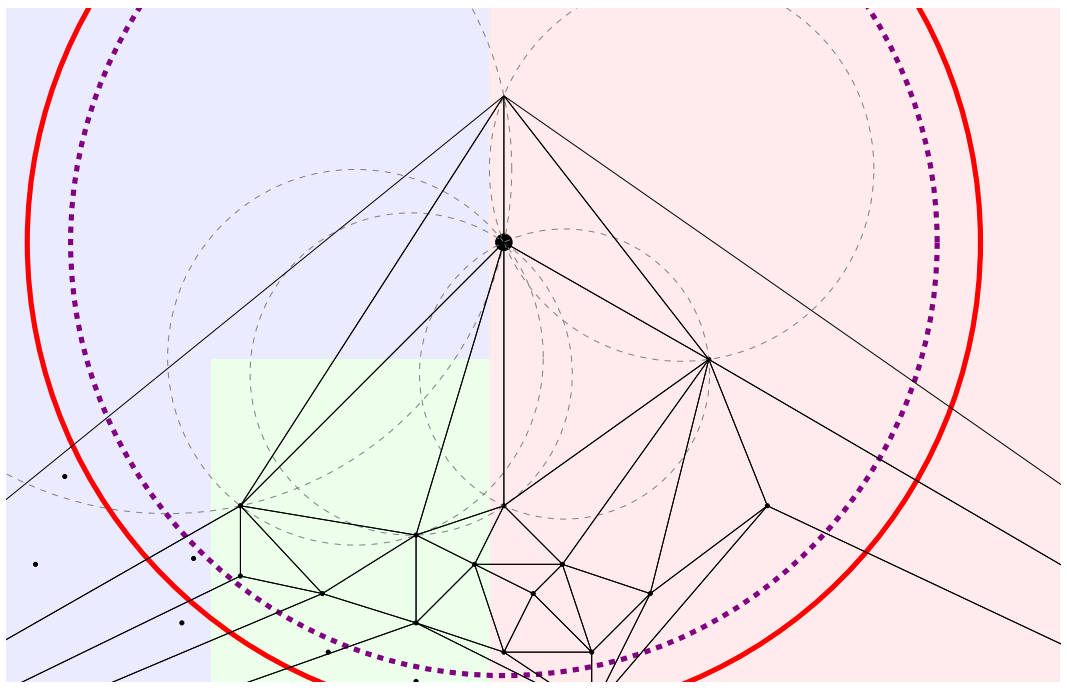

(d) In iteration 22 the testing circle brings 3 points to be added to the Delaunay triangulation. The point will now be classified as large.

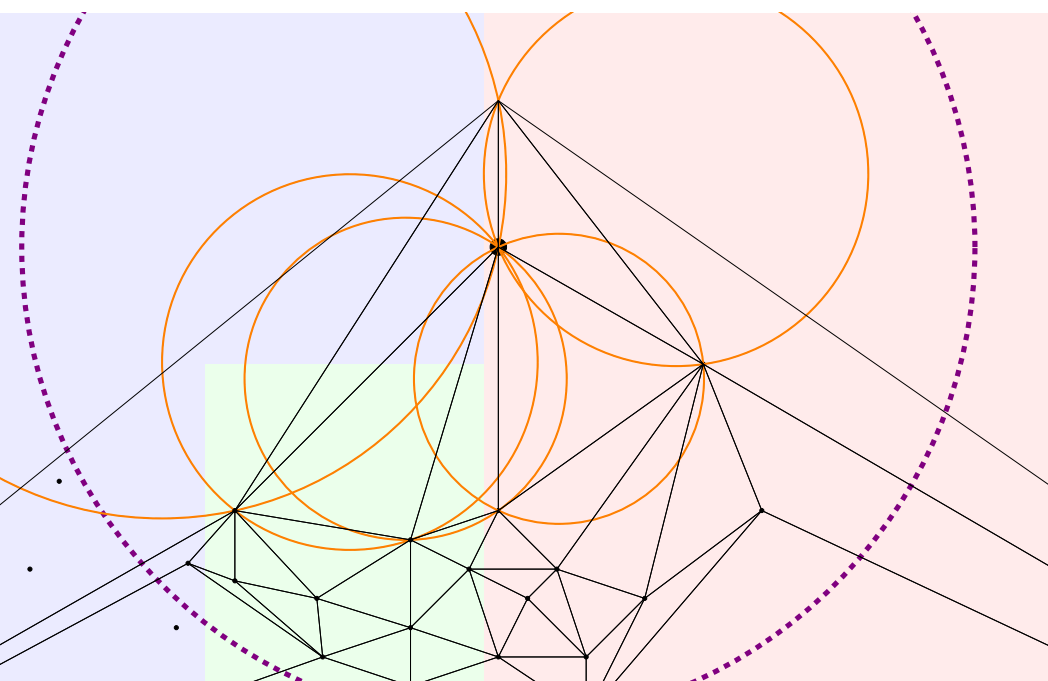

(e) Each one of the circumcircles is tested individually. One point is brought from the blue processor.

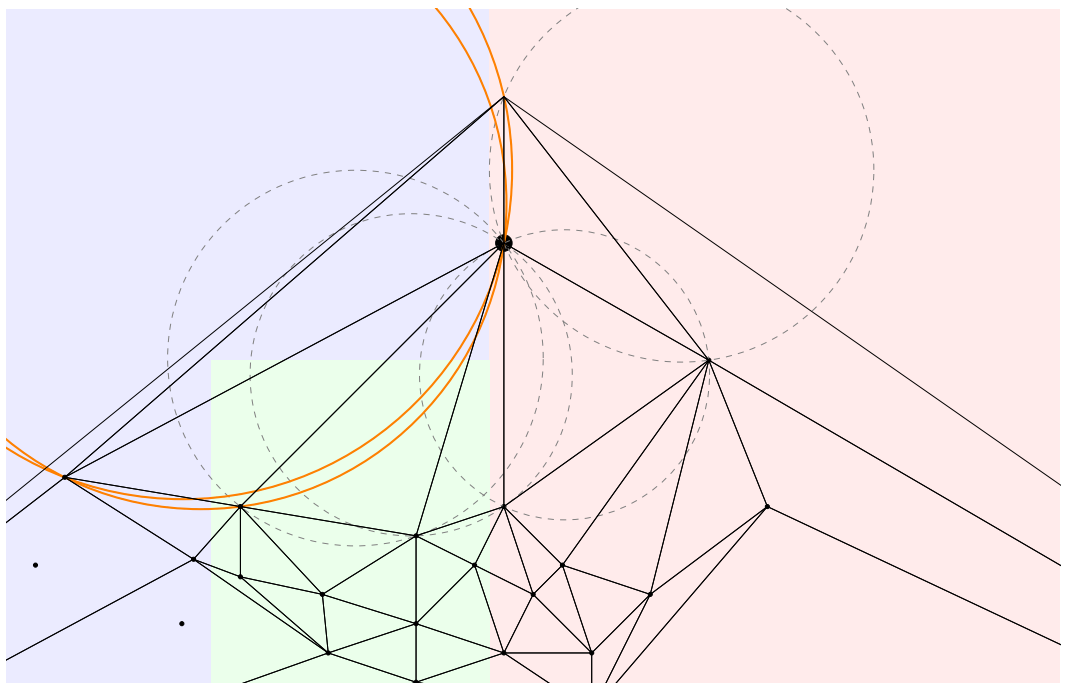

(f) The point is added to the Delaunay triangulation. Circumcircles of the new triangles created are tested now.

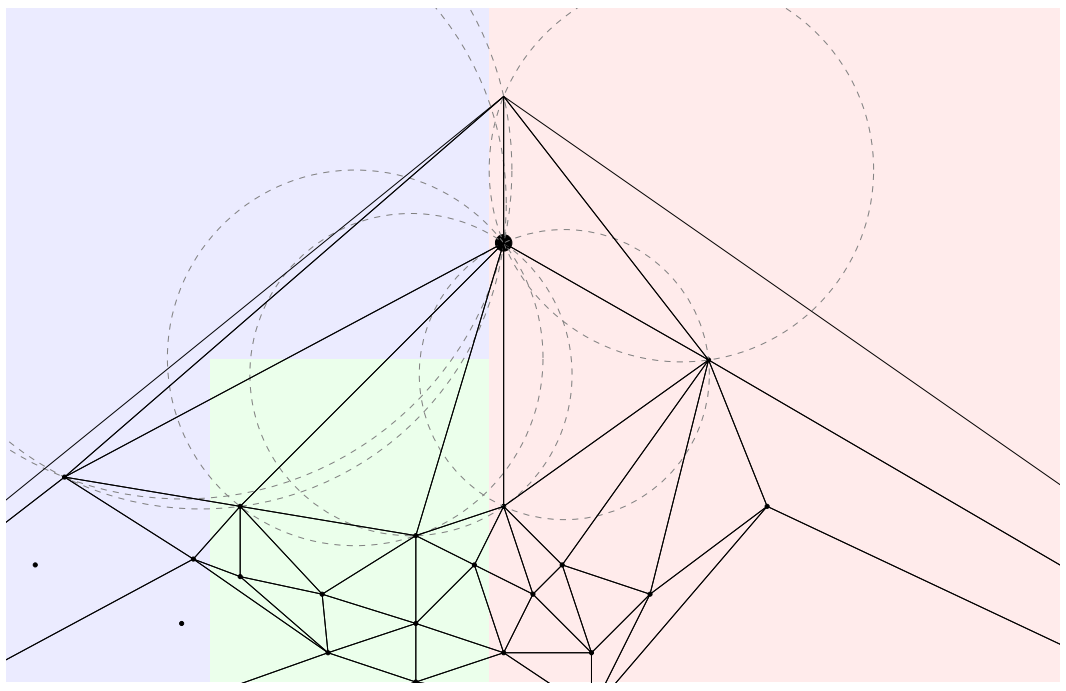

(g) The point is added to the local Delaunay triangulation. No further circles to be tested, so we are done.

**Figure 11.** A demonstration of our proposed algorithm for one point of the red processor. The background colours denote the processor affinity of points. The red circle is the testing circle (in the small phase), and the dashed purple is the previous testing circle. The dotted grey circles are the current circumcircles of the point's triangles. In the big phase, tested circumcircles will be coloured orange.

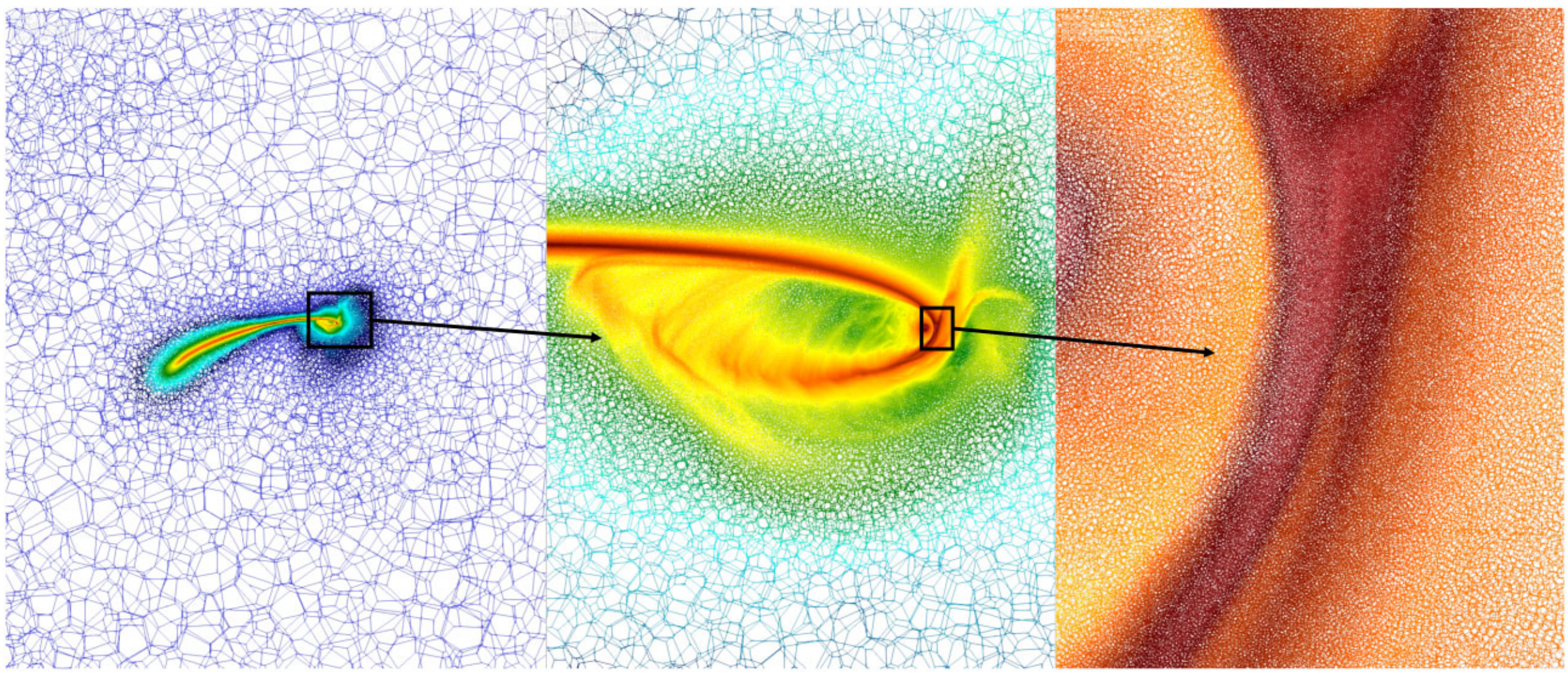

**Figure 12.** The geometry of the Voronoi cells for the astrophysical data set colour coded by their density. The middle and the right figures are inset of the figure to their left, highlighting the large dynamical range in cell sizes.

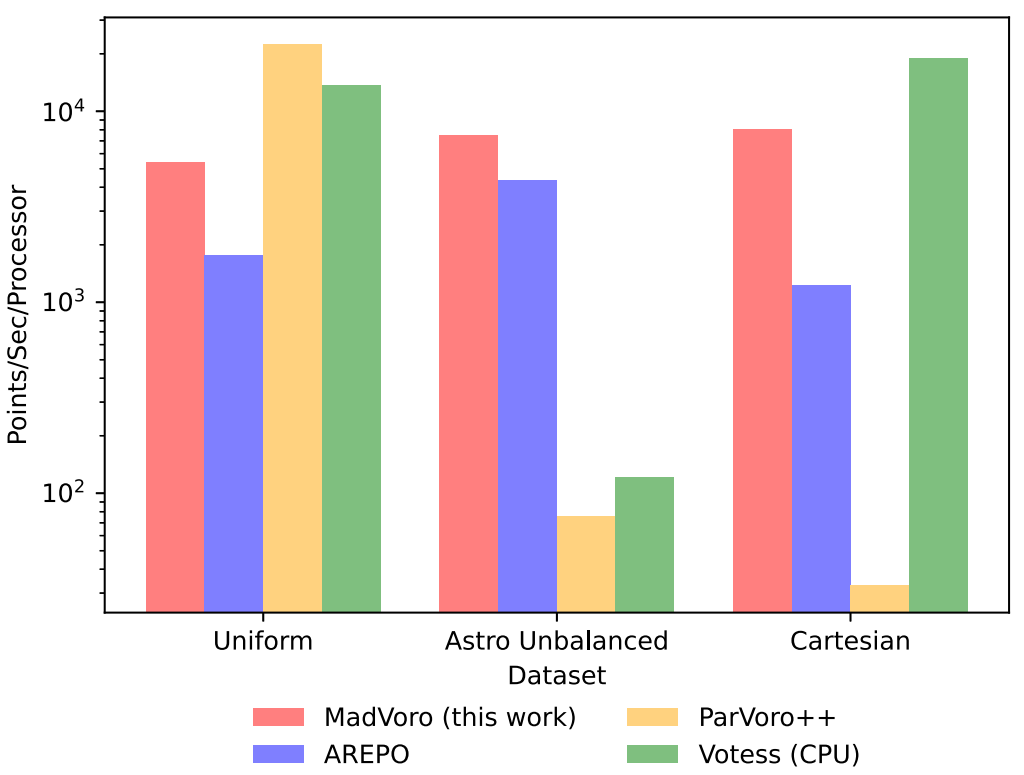

(a) The construction rate compared to different frameworks for the first build (log scale).

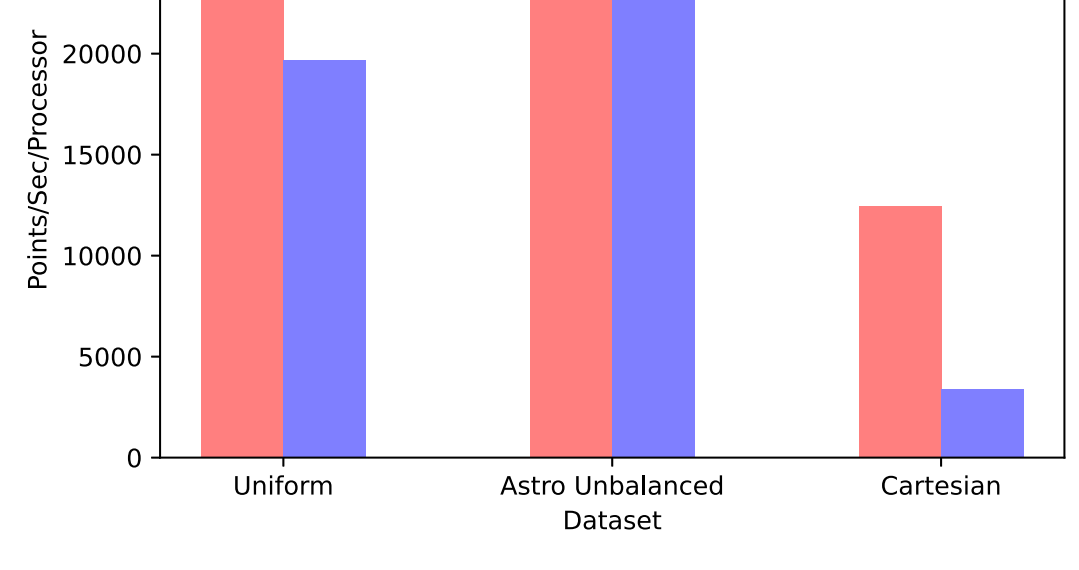

(b) The average construction rate compared of the next 5 constructions.

**Figure 13.** Results of parallel execution (16 cores).

### 7.1 Memory consumption

The average memory consumption of the construction process, tested on an execution of 128 processors, each holding 10 000 random points in the unit square ($[0, 1]^3$), is approximately 171 MB per core. On average, each point consumes 17.5KB. It is important to note that the peak memory usage accounts for both the load balancing and the Delaunay triangulation and Voronoi diagram construction processes, as well as the data structures stored for future access to the diagram's information. This includes the vertices and faces of cells, centres of mass for both cells and faces, cell volumes and face areas, neighbour relationships, and – when using parallel construction – data related to the sent and received ghost points.

### 7.2 Comparing to parallel frameworks

Here, we compare our execution to parallel frameworks, including our code and AREPO, in a single machine of 16 cores (capable of running 16 threads or MPI processes). We compare here the first build only. In PARVORO++ (Wu et al. (2023)) we determined the optimal number of blocks for partitioning by running multiple executions with varying options (powers of 2), selecting the best configuration for each data set (which, in all cases, turned out to be 512). For VOTESS (Singh et al. 2024), the points were scaled into $[0, 1]^3$, preserving the ratio of distances. We present three data sets:

(i) Random points sampled, each one, uniformly in the cube $[0, 1]^3$ (all frameworks use the same sampled points).

(ii) Astrophysical data set, taken from a simulation of a half a solar mass star being tidally disrupted by a $10^5$ $M_\odot$ SMBH. The geometry of the Voronoi cells in the midplane colour-coded by their density is shown in Fig. 12.

(iii) a Cartesian grid, dividing the unit square $[0, 1]^3$ to $100 \times 100 \times 100$.

Fig. 13(a) shows the average construction rate (points per second for a single processor) of the first build. PARVORO++ and VOTESS do not support successive builds, and therefore we compare ourselves to Springel separately in Fig. 13(b), testing the average

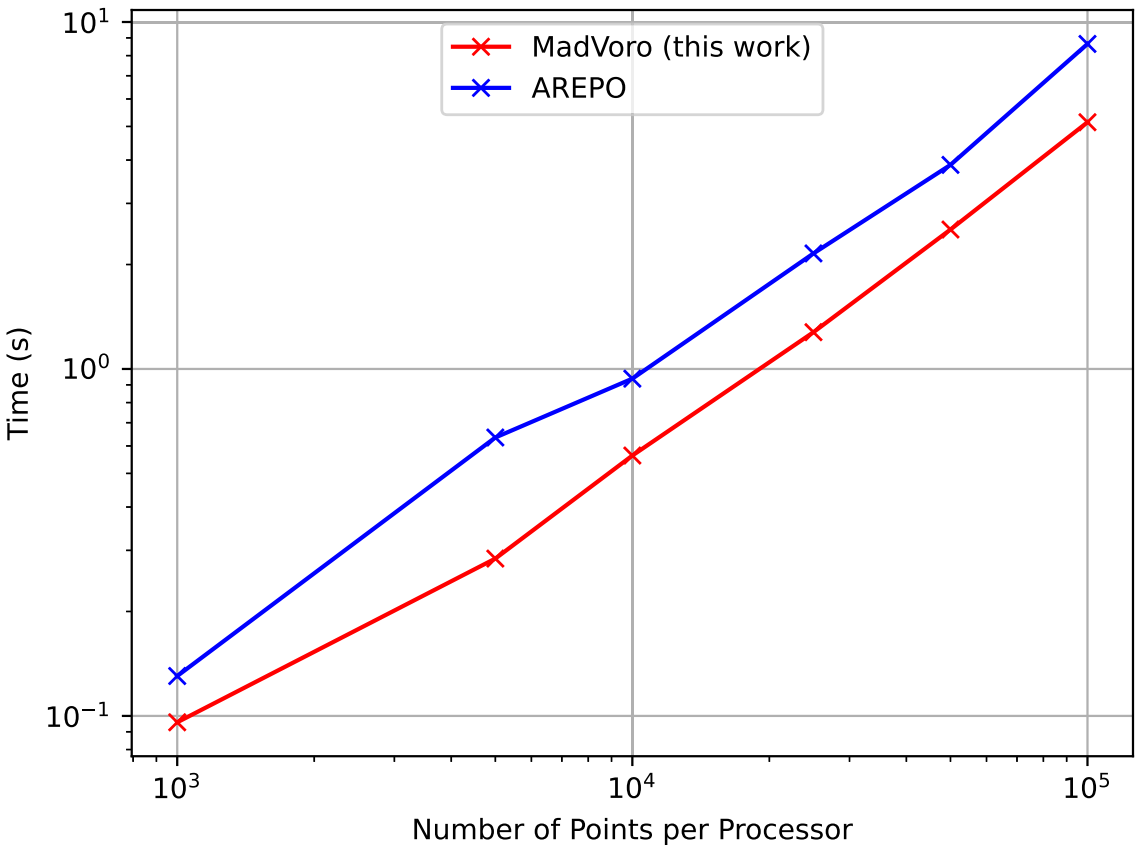

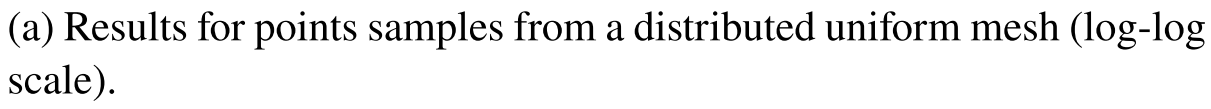
(a) Results for points samples from a distributed uniform mesh (log-log scale).

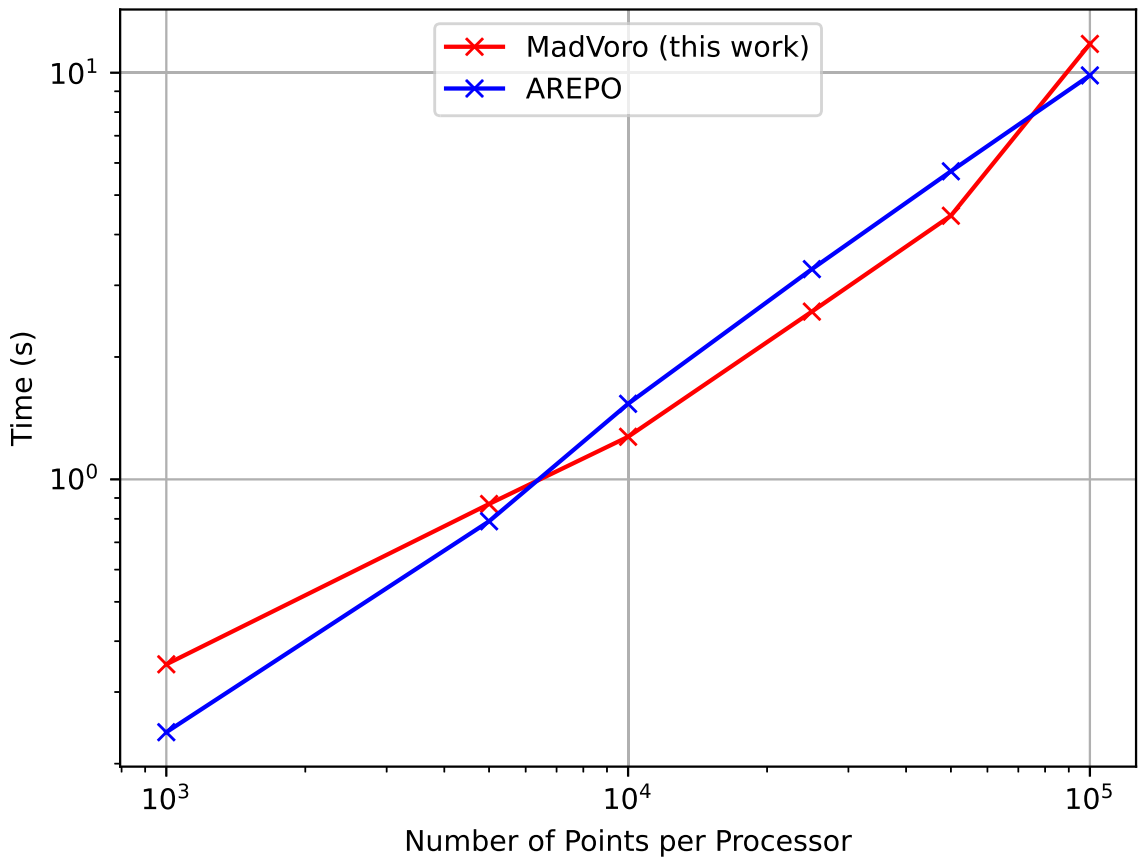

(b) Results for points located in an unbalanced star-like shape mesh, as described earlier (log-log scale).

**Figure 14.** Results of distributed execution (512 cores).

construction time of the next 5 builds. Graphs are shown in log scale.

As seen, our construction method performs exceptionally well in more complex scenarios while maintaining efficient behaviour even in a uniform setting. VOTESS excels in the Cartesian settings since its construction algorithm does not involve Delaunay triangulations (the Cartesian grid causes degeneracies and high precision calculations, consuming more time). It's important to highlight that a uniform mesh is relatively uncommon in real-world applications, where more complex point distributions are typical. One might initially expect the uniform mesh to be constructed more quickly than the unbalanced astrophysical mesh. However, profiling reveals that the difference in the core algorithm's running time between the two benchmarks is relatively minor. The primary discrepancy lies in the handling of ghost points outside the user-defined box.[11] In the uniform data set, significantly more points require mirroring compared to the astrophysical data set, where the majority of points are concentrated near the centre and fewer reside near the boundaries. As a result, inserting mirrored points into the Delaunay triangulation accounts for nearly 16 per cent of the total runtime in the uniform case, while it is negligible in the astrophysical case.

## 7.3 Comparison with AREPO

### *7.3.1 Uniform mesh*

In this data set, the points are uniformly distributed in $[0, 1]^3$. Results are shown in Fig. 14(a). As one can see, our new suggested algorithm improves all the other listed methods.

### *7.3.2 Star-like mesh*

In this data set, the points distribution is as follows: the whole space is a cube with a side length of 1 ($[0, 1]^3$). 95 per cent of the points are uniformly distributed inside the cube $[0.45, 0.5]^3$ (a cube with a side length of 0.05). The other 5 per cent are uniformly distributed in all the space. The remaining 5 per cent points may also fall in the small cube. The results are shown in Fig. 14(b).

[11] Referred to as *mirrored points* in our terminology. These points are introduced to ensure the preservation of the user-specified box geometry and are generated by reflecting boundary points across the faces of the box, thereby enforcing the presence of those faces in the final structure.

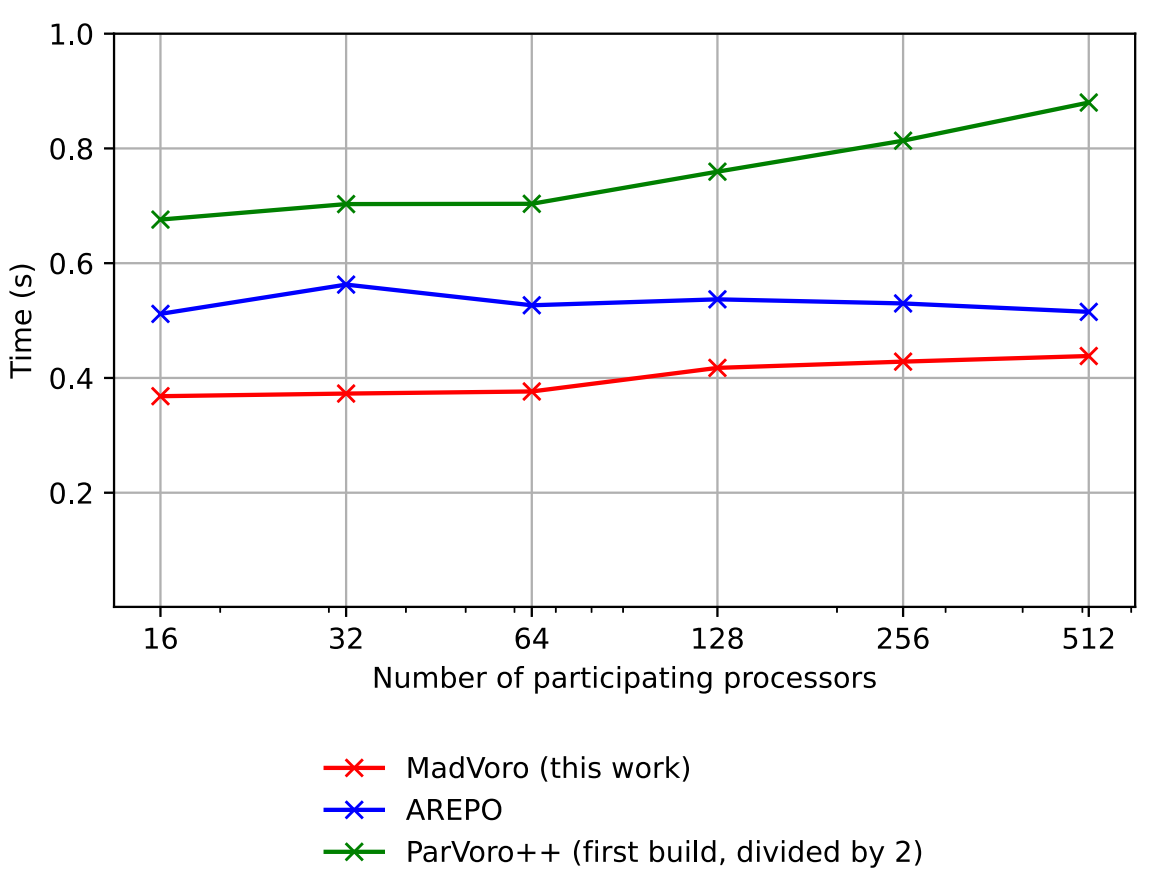

**Figure 15.** A weak scale execution of a uniform mesh. The number of points per processor is determined to be 10 000 (the $x$ axis is in log scale). Note that the curve for PARVORO++ has been scaled down by a factor of two, as its original values were too large to fit within the graph.

## 7.4 Scalability

### *7.4.1 Weak scale*

To demonstrate the weak scale of MADVORO, we test the running time of advanced builds (average of the first 10 builds) of a mesh of points uniformly sampled in $[0, 1]^2$. We compare our code to Springel's implementation (AREPO), when conserving the number of points per processor to be 10 000, and changing the number of participating processors from 8 to 512. Results are shown in Fig. 15, where the $x$ axis, representing the number of processors, is in log scale. It is clear that although there is a room for improvement, MADVORO preserves weak scale good enough.

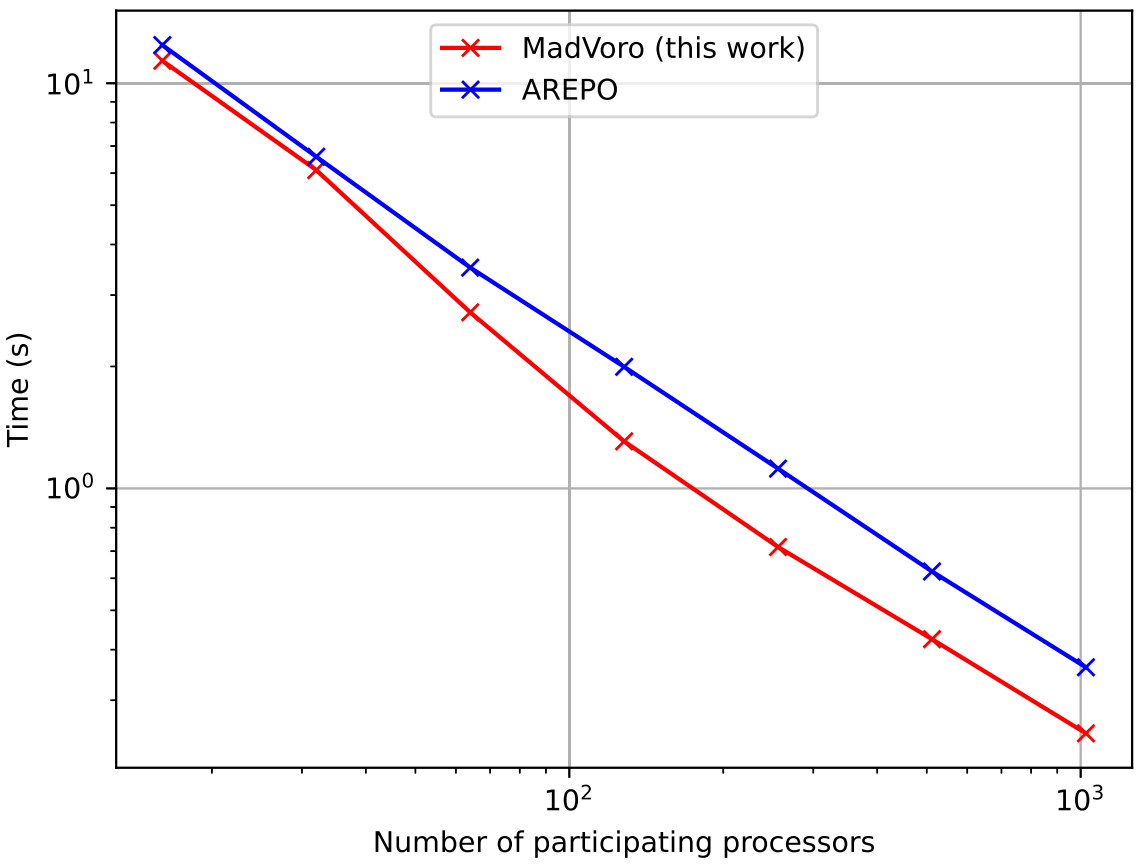

**Figure 16.** A strong scale execution on the astrophysical data set (log–log scale).

### 7.4.2 *Strong scaling*

To test the strong scaling, we used the same astrophysical data set from Section 7.2. The total number of points is determined to 5127679, and the number of processors varies, from 8 to 1024. Again, the measured times represent the average of the first 10 constructions following the initial one. The ideal scalability, given $P$ points in total and $N$ processors, is $N/P$. Since the graph is displayed in log-log scale, a linear curve is expected. MADVORO consistently outperforms AREPO in construction time, with both methods demonstrating excellent strong scaling. Results are shown in Fig. 16.

## 8 CONCLUSION AND FUTURE WORK

We discussed Voronoi diagrams and their duality with Delaunay triangulations, focusing on the construction problem of a distributed Delaunay triangulation. We introduced Springel's algorithms (Springel 2010) and highlighted their limitations in building an unbalanced mesh. By merging both algorithms and improving their bottlenecks, we created a new, more efficient algorithm. Our assessments, based on several benchmarks, demonstrated that the new algorithm either improved upon or performed at least as well as all other assessed algorithms.

Delaunay triangulation in a distributed memory parallel system requires careful load balancing to ensure efficiency. Load balancing plays a crucial role by facilitating the construction process, as processors are more familiar with the communication patterns and the mesh decomposition that define it. We explored curve-based balancing, using the Hilbert curve as an example to demonstrate its advantages.

Future work will focus on enhancing the curve-based technique and exploring its generalization to other shapes and curves. A method we introduced previously, called *kernelization* (Mizrachi et al. 2024), could be one avenue for this exploration. Additionally, there may be further improvements to be made to the construction algorithm itself. Specifically, the current approach – computing the Voronoi tessellation using Delaunay triangulation duality, followed by the flipping algorithm for Delaunay triangulation – may not be the most optimal. Moreover, we are considering the potential limitations of the communication paradigm currently in use. While we have implemented a query-based communication approach, it might be worthwhile to investigate alternative communication paradigms that could offer better performance or scalability in the presented scheme.

## ACKNOWLEDGEMENTS

The principal author thanks Rainer Weinberger for his assistance in running AREPO. We also express our gratitude to the computational physics department at the Racah Institute of Physics, particularly Omri Reved, for their helpful discussions.

## CONFLICT OF INTEREST

Authors declare no conflict of interest.

## DATA AVAILABILITY

Our construction framework is available to all as an open-source project in Github. Installation and execution instructions can be found in the README file.

## REFERENCES

Alliez P., Delage C., Karavelas M. I., Pion S., Teillaud M., Yvinec M., 2010, Research Report, Delaunay Tessellations and Voronoi Diagrams in CGAL, INRIA Sophia Antipolis – Méditerranée; University of Cretea, available at: https://inria.hal.science/hal-01421021
Bauman K. E., 2006, Math. Notes, 80, 609
Bokhare A., Metkewar P., 2019, Int. J. Recent Technol. Eng., 8, 775
Borrell R., Cajas J., Mira D., Taha A., Koric S., Vaz`quez M., Houzeaux G., 2018, Comput. Fluids, 173, 264
Chen J., Saad Y., Zhang Z., 2022, SeMA J., 79, 187
Cignoni P., Montani C., Scopigno R., 1998, Comput. Aided Des., 30, 333
Duffell P. C., MacFadyen A. I., 2011, ApJS, 197, 15
Edelsbrunner H., Shah N. R., 1996, Algorithmica, 15, 223
Elshakhs Y. S., Deliparaschos K. M., Charalambous T., Oliva G., Zolotas A., 2024, IEEE Access, 12, 12562
Filipiak M., 2013, Mesh Reordering in Fluidity using Hilbert Space-Filling Curves. EPCC, University of Edinburgh
Gonzalez R. E., 2016, Astron. Comput., 17, 80
Gotsman C., Lindenbaum M., 1996, IEEE Trans. Image Process., 5, 794
Harlacher D., Klimach H., Roller S., Siebert C., Wolf F., 2012, Dynamic Load Balancing for Unstructured Meshes on Space-Filling Curves. p. 1661
Karypis G., Kumar V., 1998, SIAM J. Sci. Comput., 20, 359
Ledoux H., 2007, in 4th International Symposium on Voronoi Diagrams in Science and Engineering (ISVD 2007). p. 117
Lo S., 2012, Comput. Meth. Appl. Mech. Eng., 237–240, 88
Mizrachi M., Steinberg E., Raveh B., 2024, Parallel Constructing of Voronoi Diagrams and Delaunay Triangulations for Distributed Memory Physics Simulations. The Hebrew University of Jerusalem
Moon B., Jagadish H., Faloutsos C., Saltz J., 2001, IEEE Trans. Knowl. Data Eng., 13, 124
Morozov D., Peterka T., 2016, in SC'16: Proceedings of the International Conference for High Performance Computing, Networking, Storage and Analysis
Peterka T., Morozov D., Phillips C., 2014, SC '14: Proceedings of the International Conference for High Performance Computing, Networking, Storage and Analysis. p. 997
Pothen A., Simon H. D., Liou K.-P., 1990, SIAM J. Matrix Anal. Appl., 11, 430
Potter D., Stadel J., Teyssier R., 2017, Computat. Astrophys. Cosmol., 4, 2
Rycroft C. H., 2009, Chaos: An Interdiscipl. J. Nonlin. Sci., 19, 041111
Sasidharan A., Dennis J. M., Snir M., 2015, in IEEE 17th International Conference on High Performance Computing and Communications (HPCC), IEEE 7th International Symposium on Cyberspace Safety and Security (CSS) and IEEE 12th International Conf on Embedded Software and Systems (ICESS). IEEE Computer Society, Los Alamitos, CA, p. 875
Shewchuk J. R., 1997, Discrete Comput. Geometry, 18, 305
Singh S. D., Byrohl C., Nelson D., 2024, preprint (arXiv:2412.04514)

Springel V., 2010, MNRAS, 401, 791
Steinberg E., Yalinewich A., Sari R., Duffell P., 2015, ApJS, 216, 14
Wadsley J., Stadel J., Quinn T., 2004, New Astron., 9, 137
Walshaw C., Cross M., 2007, in Magoulès F., ed., Computational & Technology Resources. Saxe-Coburg Publications, Stirlingshire, UK, p. 27
Watson D., 1993, Comput. Geosci., 19, 1209
Wu G., Tian H., Lu G., Wang W., 2023, Parallel Comput., 115, 102995
Yalinewich A., Steinberg E., Sari R., 2015, ApJS, 216, 35